\documentclass[11pt]{amsart}

\usepackage{amscd}
\usepackage{amsthm}
\usepackage{amssymb}
\usepackage{hyperref}

\newtheorem{thm}{Theorem}[section]

\newtheorem{propn}[thm]{Proposition}

\newtheorem{conj}[thm]{Conjecture}

\theoremstyle{definition}
\newtheorem{dfn}[thm]{Definition}
\newtheorem{rem}[thm]{Remark}
\newtheorem{examp}[thm]{Example}
\newtheorem{question}[thm]{Question}

\newcommand{\Cat}{{\textbf{C}}}
\newcommand{\smproj}{{\textbf{SmProj}}_k}
\newcommand{\sch}{{\textbf{Sch}}_k}
\newcommand{\quasiproj}{{\textbf{Var}}_k}
\newcommand{\relvar}{{\textbf{Var}}}
\newcommand{\modu}{{\textbf{Mod}}_R}
\newcommand{\vect}{{\textbf{Vect}}_k}
\newcommand{\grvect}{{\textbf{GrVect}}}
\newcommand{\sets}{{\textbf{Set}}}
\newcommand{\repG}{{\textbf{Rep}}_k (G)}
\newcommand{\repk}{{\textbf{Rep}}_k}
\newcommand{\etalecov}{{\textbf{\'EtCov}}}
\newcommand{\abelian}{{\textbf{Ab}}}
\newcommand{\obj}{{\textrm{Obj} \ }}
\newcommand{\abgps}{{\textbf{AbGp}}}

\renewcommand{\dim}{{\rm dim}}

\newcommand{\limproj}{\mathop{\varprojlim}\limits}

\newcommand{\cycle}{C^{i}_\sim }
\newcommand{\ichow}{\textup{CH}^i}
\newcommand{\cor}{\textup{Corr}}

\newcommand{\lef}{\mathbb{L}}
\newcommand{\torus}{\mathbb{T}}
\newcommand{\point}{\mathbb{I}}
\newcommand{\rat}{\sim_{\textup{rat}}}
\newcommand{\num}{\sim_{\textup{num}}}
\newcommand{\homo}{\sim_{\textup{hom}}}

\newcommand{\ra}{\rightarrow}           
\newcommand{\lra}{\longrightarrow}
\newcommand{\hra}{\hookrightarrow}

\newcommand{\Z}{{\mathbb{Z}}}
\newcommand{\R}{{\mathbb{R}}}
\newcommand{\C}{{\mathbb{C}}}
\newcommand{\Q}{{\mathbb{Q}}}
\newcommand{\N}{{\mathbb{N}}}
\newcommand{\F}{{\mathbb{F}}}
\newcommand{\Gm}{{\mathbb{G}}_m}

\newcommand{\Gal}{{\textup{Gal}}}
\newcommand{\Hom}{{\textup{Hom}}}
\newcommand{\End}{{\textup{End} \ }}
\newcommand{\Ext}{{\textup{Ext}}}
\newcommand{\Tor}{{\textup{Tor}}}

\newcommand{\ihom}{\underline{\Hom}}
\newcommand{\Aut}{{\textup{Aut}}}
\newcommand{\TenAut}{{\textup{Aut}}^{\otimes}}
\newcommand{\Spec}{{\textup{Spec} \ }}

\newcommand{\id}{{\textrm{id}}}
\newcommand{\tr}{{\textrm{tr}} \ }
\newcommand{\ev}{{\textrm{ev}} \ }

\newcommand{\gr}{{\textrm{gr}}}
\newcommand{\mhom}{{\textbf{Mot}}_{{\textrm{hom}}}}
\newcommand{\mchow}{{\textbf{Mot}}_{{\textrm{rat}}}}
\newcommand{\mnum}{{\textbf{Mot}}_{{\textrm{num}}}}
\newcommand{\mot}{{\textbf{Mot}}}

\newcommand{\mtmot}{{\textbf{MTMot}}}
\newcommand{\mhs}{{\textbf{MHodge}}}
\newcommand{\hs}{{\textbf{Hodge}}}
\newcommand{\real}{{\textbf{real}}}

\newcommand{\A}{{\mathbb{A}}}
\newcommand{\Proj}{{\mathbb{P}}}

\newcommand{\blow}{{\textup{Bl}}} 
\newcommand{\scrl}{{\mathcal{L}}}
\newcommand{\scrx}{{\mathcal{X}}}
\newcommand{\ord}{{\textup{ord}}}
\newcommand{\sing}{{\textup{Sing} \ }}

\begin{document}
\title{Motives: an introductory survey for physicists}
\author{Abhijnan Rej \\ (with an appendix by Matilde Marcolli)}
\email{abhijnan.rej@gmail.com}
\urladdr{http://abhijnan.rej.googlepages.com}
\thanks{Part of this work was done while the author was employed by the Clay Mathematics Institute as a Junior Research Scholar.}
\begin{abstract}
We survey certain accessible aspects of Grothendieck's theory of motives in arithmetic algebraic geometry for mathematical physicists, focussing on areas that have recently found applications in quantum field theory. An appendix (by Matilde Marcolli) sketches further connections between motivic theory and theoretical physics.
\end{abstract}
\maketitle

\begin{verse}
\emph{post hoc, ergo ante hoc}\\
-- Umberto Eco, \emph{Interpretation and overinterpretation}
\end{verse}
\vspace{5mm}

\setcounter{tocdepth}{1} 
\tableofcontents

\section{Introduction}

This survey paper is based on lectures given by the author at Boston University, the Max Planck Institute in Bonn, at Durham University and at the Indian Statistical Institute and the SN Bose National Center for the Basic Sciences in Kolkata and the Indian Institute of Technology in Mumbai. The purpose of these introductory notes are to familiarize an audience of physicists with some of the algebraic and algebro-geometric background upon which Grothendieck's theory of motives of algebraic varieties relies. There have been many recent developments in the interactions between high energy physics and motives, mostly within the framework of perturbative quantum field theory and the evaluation of Feynman diagrams as periods of algebraic varieties, though motives are beginning to play an important role in other branched of theoretical physics, such as string theory, especially through the recent interactions with the Langlands program, and through the theory of BPS states. We focus here mostly on the quantum field theoretic applications when we need to outline examples that are of relevance to physicists. The appendix to the paper, written by Matilde Marcolli, sketches several examples of how the main ideas involved in the theory of motives, algebraic cycles, periods, Hodge structures, K-theory, have in fact been already involved in many different ways in theoretical physics, from condensed matter physics to mirror symmetry. Most of the paper focuses on the mathematical background. 
\par
We describe the Grothendieck ring of varieties and its properties, since this is where most of the explicit computations of motives associated to Feynman integrals are taking place. We then discuss the Tannakian formalism, because of the important role that Tannakian categories and their Galois groups play in the theory of perturbative renormalization after the work of Connes--Marcolli. We then describe the background cohomological notions underlying the construction of the categories of pure motives, namely the notion of Weil cohomology, and the crucial role of algebraic cycles in the theory of motives. It is in fact mixed motives and not the easier pure motives are involved in the application to quantum field theory, due to the fact that the projective hypersurfaces associated to Feynman graphs are highly singular, as well as to the fact that relative cohomologies are involved since the integration computing the period computation that gives the Feynman integral is defined by an integration over a domain with boundary. We will not cover in this survey the construction of the category of mixed motives, as this is a technically very challenging subject, which is beyond what we are able to cover in this introduction. However, the most important thing to keep in mind about mixed motives is that they form {\em triangulated} categories, rather than abelian categories, except in very special cases like mixed Tate motives over a number field, where it is known that an abelian category can be constructed out of the triangulated category from a procedure known as the heart of a t-structure. For our purposes here, we will review, as an introduction to the topic of mixed motives, some notions about triangulated categories, Bloch--Ogus cohomologies, and mixed Hodge structures. We end by reviewing some the notion of motivic zeta function and some facts about motivic integration. Although this last topic has not yet found direct applications to quantum field theory, there are indications that it may come to play a role in the subject.
\par
A word about the references: the list is much longer than the list of works cited in the main body of the text. This is deliberate since the list is meant to also serve as a guide to futher reading.

\subsection*{Acknowledgements} These notes, as noted in the introduction, form a part of my lectures on motives at various institutions in the US, Germany, the UK and India during 2005-2009, in particular, my talk at the Bonn workshop on renormalization in December 2006. I thank the organizers (Ebrahimi-Fard, van Suijlekom and Marcolli) for a stimulating meeting and for their infinite patience in waiting for this written contribution towards the proceedings of that meeting.

\section{The Grothendieck ring}\label{groth}
\subsection{Definition of the Grothendieck groups and rings}
Recall the definition of the usual Grothendieck group of vector bundles on a smooth variety $X$:
\begin{dfn}[exercise 6.10 of \cite{hart}] \label{Ktop}
$K_0(X)$ is defined as the quotient of the free abelian group generated by all vector bundles (= locally free sheaves) on $X$  by the subgroup generated by expressions $\mathcal{F} - \mathcal{F}' - \mathcal{F}''$ whenever there is an exact sequence of vector bundles $0 \ra \mathcal{F}' \ra \mathcal{F} \ra \mathcal{F}'' \ra 0$. The group $K_0(X)$ can be given a ring structure via the tensor product.
\end{dfn}
We remark that isomorphism classes of vector bundles form an abelian monoid under direct sum; in fact, Grothendieck groups can be defined in a much more general way because of the following universal property:

\begin{propn}\label{Kuniv}
Let $M$ be an abelian monoid. There exists an abelian group $K(M)$ and $\gamma: M \ra K(M)$ a monoid homomorphism satisfying the following universal property: If $f: M \ra A$ is a homomorphism into an abelian group $A$, then there exists a \emph{unique} homomorphism of abelian groups $f_{*}: K(M) \ra A$ such that the following diagram commutes:
\begin{equation*}\begin{CD}
M @>\gamma >> K(M) \\
@V{f}VV           @VV{f_{*}}V\\
A @= A
\end{CD} \end{equation*}
\end{propn}
The proof of proposition \ref{Kuniv} is very simple: construct a free abelian group $F$ generated by $M$ and let $[x]$ be a generator of $F$ corresponding to the element $x \in M$. Denote by $B$ the subgroup of $F$ generated by elements of the form $[x+y] - [x] - [y]$ and set $K(M)$ to be the quotient of $F$ by $B$. Letting $\gamma$ to be the injection of $M$ into $F$ and composing with the canonical map $F \ra F /B$ shows that $\gamma$ satifies the universal property. 
\par
One can show that projective modules over a ring $A$ give rise to a Grothendieck group denoted as $K(A)$ (p. 138 of \cite{lang}). This is simply done by noting that isomorphism classes of (finite) projective $A$-modules form a monoid (again under the operation of direct sum) and taking the subgroup $B$ to generated by elements of the form $[P \oplus Q] - [P] -[Q]$ for finite projective modules $P$ and $Q$. We can refine $K(A)$ by imposing the following equivalence: $P$ is equivalent to $Q$ $\iff$ there exists free modules $F, F'$ such that $[P \oplus F] = [Q \oplus F']$. Taking the quotient of $K(A)$ with respect to this equivalence relation gives us $K_0(A)$. 
\par
The explicit determination of $K_0$ (and its ``higher dimensional" analogues) for a given ring $A$ is the rich subject of algebraic $K$-theory. (A standard reference for algebraic $K$-theory is \cite{rosenberg}; see also the work-in-progress \cite{weibel}.)  For the simplest case when $A$ is a field, we note that $K_0(A) \simeq \Z$; this immediately follows from the definition and the trivial fact that modules over a field are vectors spaces. As a non-trivial example, we have the following
\begin{examp}[example 2.1.4 of \cite{weibel}]
Let $A$ be a semisimple ring with $n$ simple modules. Then $K_0(A) \simeq \mathbb{Z}^n$.
\end{examp}
\par
Through the theorem of Serre-Swan which demonstrates that the categories of vector bundles and finite projective modules are equivalent, we see that definition \ref{Ktop} arise very naturally from proposition \ref{Kuniv}.  Furthermore, the generality of proposition \ref{Kuniv} enables us construct Grothendieck groups and rings of (isomorphism classes of) other objects as well. 
\par
Let $\quasiproj$ the category of quasi-projective varieties over a field $k$. Let $X$ be an object of $\quasiproj$ and $Y \hookrightarrow X$ a closed subvariety. Central to our purposes would be the following
\begin{dfn} \label{Kvar} 
The Grothendieck ring $K(\quasiproj)$ is the quotient of the free abelian group of isomorphism classes of objects in $\quasiproj$ by the subgroup generated the expressions $[X] - [Y] - [X \setminus Y]$. The ring structure given by fiber product: $[X]\cdot [Y] := [X \times_k Y]$.
\end{dfn}
In fact one can generalize definition \ref{Kvar} to define the Grothendieck ring of a \emph{symmetric monodial category}. Recall the following
\begin{dfn}[defintion 5.1 of chapter II of \cite{weibel}] \label{symmon}
A category $\Cat$ is called \emph{symmetric monoidal} if there is a functor $\otimes : \Cat \ra \Cat$ and a distinguished $\mathbb{I} \in \obj(\Cat)$ such that the following are isomorphisms for all $S, T, U  \in \obj(\Cat)$:
\begin{eqnarray*}
                                   \mathbb{I} \otimes S & \cong & S  \\
                                   S \otimes \mathbb{I} & \cong & S\\ 
                                   S \otimes (T \otimes U) & \cong & (S \otimes T) \otimes U \\
                                   S \otimes T & \cong & T \otimes S.
\end{eqnarray*}
\end{dfn}
Furthermore one requires the isomorphisms above to be \emph{coherent}, a technical condition that guarantees that one can write expressions like $S_1 \otimes \ldots \otimes S_n$ without paranthesis without ambiguity\footnote{Note that we will sometimes call a symmetric monodial category an ``ACU $\otimes$-category" and refer to the functor $\otimes$ as the \emph{tensor functor}.}. (See any book on category theory for the precise definitions.)
\par
\begin{rem}
The category $\quasiproj$ is symmetric monodial with the functor $\otimes$ given by fiber product of varieties and the unit object $\mathbb{I}$ given by $\mathbb{A}^0 = \textup{point}$. 
\end{rem}
An important example of a symmetric monoidal category is furnished by the category of finite dimensional complex representations of a finite group $G$ (with the morphisms given by intertwinners between representations), denoted as $\textbf{Rep}_\C(G)$.
\begin{examp}[example 5.2.3 of \cite{weibel}] \label{repring}
$\textbf{Rep}_\C(G)$ is symmetric monodial under direct sums of representations. Furthermore, $K_0(\textbf{Rep}_\C(G)) \simeq R(G)$ generated by irreducible representations and where $R(G)$ is the representation ring of $G$.
\end{examp} 
Note that for more general $G$ (say an affine group scheme over an arbitrary base), the category $\textbf{Rep}_\C(G)$ has much more structure that being merely symmetric monodial, namely it is a Tannakian category. We will turn to this rich subject in section \ref{tannaka}.
\subsection{The Grothendieck ring of varieties $K(\quasiproj)$} Recall definition \ref{Kvar}. The Grothendieck ring $K(\quasiproj)$ has the following properties (cf. the review \cite{neeraja}):
\begin{enumerate}
\item{If $X$ is a variety and $U,V$ locally closed subvarieties in $X$, then $$[U \cup V] + [U \cap V] = [U] +[V].$$}
\item{If $X$ is the disjoint union of locally closed subvarieties $X_1, \ldots, X_n$, then $[X] = \sum_{i=1}^n [X_i]$.}
\item{Let $C$ be a constructible subset of a variety $X$ (that is, $C$ is the disjoint union of locally closed subsets of $X$.) Then $C$ has a class in $K(\quasiproj)$.}
\end{enumerate}
In section 6 of \cite{matilde}, Marcolli proposes a category of Feynman motivic sheaves which involves viewing the Kirchhoff polynomial as a morphism
$\Psi_\Gamma: \A^{n} \setminus X_\Gamma \ra \Gm$ where $X_\Gamma$ is viewed as an \emph{affine} hypersurface obtained by setting $\Psi_\Gamma = 0$. (Here $n$ is the number of edges of $\Gamma$ and $\Gm$ is the multiplicative group. ) As a setting for such relative questions, we introduce the following:

\begin{dfn}[Bittner, 2.1.1 of \cite{neeraja}]\label{Krel}
Let $S$ be a variety over $k$. The ring $K(\relvar_S)$ is the free abelian group generated by isomorphism classes $[X]_S$ (where $X$ is a variety over $S$) modulo the relation $[X]_S = [X \setminus Y]_S + [Y]_S$ where $Y \subset X$ is a closed subvariety. The ring structure is induced by the fiber product of varieties.
\end{dfn}
\begin{rem}
We note the following properties of $K(\relvar_S)$:
\begin{enumerate}
\item{$K(\relvar_S)$ is a $K(\quasiproj)$-module.}
\item{There is an bilinear associative exterior product 
$$K(\relvar_S) \times K(\relvar_T) \stackrel{\boxtimes}{\lra} K(\relvar_{S \times_k T}).$$}
\item{Let $f: S \ra S'$ be a morphism of varieties. It induces $f_!: K(\relvar_S) \ra K(\relvar_{S'})$ and $f^{*}: K(\relvar_{S'}) \ra K(\relvar_S)$ that are functorial with respect to $\boxtimes$: let $g: T \ra T'$ be another morphism of varieties. Then ${(f \times g)}_!(A \boxtimes B) = f_!(A) \boxtimes g_!(B)$ and ${(f \times g)}^{*}(A \boxtimes B) = f^{*}(A) \boxtimes f^{*}(B)$.}
\end{enumerate} 
\end{rem}
Note that in case of $S = \Spec k$, the definition \ref{Krel} is the same as definition \ref{Kvar}.
\begin{thm}[\cite{bittner}, from \cite{neeraja}]\label{Ksmchar}
Let $k$ a field of characteristic zero. Then $K(\quasiproj)$ is generated by smooth varieties. 
\end{thm}
The idea behind the proof is this (proof of proposition 2.1.2 of \cite{neeraja}): Set $d := \dim X$ and let $X \hra X'$ for a complete variety $X'$. Writing $[X] + [Z] = [X']$ for some $Z$ with $\dim Z \leq d-1$ and using Hironaka's theorem, we get $[\overline{X'}] = [X'] - ([C]-[E])$ where $C$ is the smooth center of the blowup $\overline{X'}$ and $E$ its exceptional divisor with $\dim C, \dim E \leq d-1$. We can write any arbitrary $X$ as the disjoint union of smooth varieties in this way by inducting on $d$.
\par
In $K(\quasiproj)$, there are two ``distinguished" classes, the class  of a point $[\mathbb{A}^0]$ and the class of the affine line $\mathbb{A}^1 = \Spec k[x]$ denoted as $\lef$. The standard cell decomposition of the projective space in terms of flags (see p.194 of \cite{grifharr}) can be lifted to $K(\quasiproj)$ in terms of these classes:
\begin{eqnarray*}
[\mathbb{P}^n] =  1 + \sum_{i =0}^{n-1} \lef^i & = & \frac{1-\lef^{n+1}}{1-\lef}\\
& = & \frac{(1+\torus)^{n+1}-1}{\torus}
\end{eqnarray*}
where $\torus := [\A^1] - [\A^0]$ the class of the torus ${\mathbb{G}}_m$. 
\par 
In order to prove this decomposition, note that $[\mathbb{P}^1_k \setminus \A^1_k] = [\mathbb{P}^1_k] - [\A^1_k]$ since $\A^1_k \subset \mathbb{P}^1_k$ is open with $\infty$ as the compliment. Induction gives the result since $\mathbb{P}^{n+1}_k \setminus \A^{n+1}_k \simeq \mathbb{P}^n_k$.
\par
One can formally invert $\lef$ in $K(\quasiproj)$ to get the \emph{Tate motive} denoted as $\Q(1)$.
By $\Q(n)$ one means $\Q(1) \cdot \cdot \cdot \Q(1)$ ($n$-times) and $n \in \Z$ is called the \emph{twist} of the Tate motive. One also sets as a matter of notation (as of now!) that $\Q(-1) := \lef$. 
\par
A rather useful fact in the context of blowups of graph hypersurfaces is the following
\begin{propn}[\cite{neeraja}]
Let $f: X \ra Y$ be a proper morphism of smooth varieties which is a blowup with the smooth center $Z \subset Y$ of codimension $d$. Then
$$[f^{-1}(Z)] = [Z] [{\mathbb{P}}^{d-1}].$$
\end{propn}
As an example of decomposition of varieties into Lefschetz motives, we take a family of hypersurfaces in $\mathbb{P}^{n-1}$
$$X_{\Gamma_n} = \{ (t_1, \ldots, t_n) | \Psi_{\Gamma_n} = t_1 \cdots t_n \Big(\frac{1}{t_1} + \cdots + \frac{1}{t_n} \Big) = 0 \}.$$
The polynomial $\Psi_{\Gamma_n}$ is the Kirchoff polynomial attached to a graph with two vertices and $n$ number of parallel edges between them (the \emph{banana graphs}); it arises in quantum field theory in the denominator of the Schwinger parametrized integral attached to the same graph. Then we have
\begin{examp}[Aluffi-Marcolli, theorem 3.10 of \cite{paomat}] \label{paomat} The class associated to the banana graph hypersurfaces $X_{\Gamma_n}$ is given by
\begin{eqnarray*}
[X_{\Gamma_n}] & = & \frac{{(1+ \mathbb{T})}^n - 1}{\mathbb{T}} - \frac{\mathbb{T}^n - {(-1)}^n}{\mathbb{T} + 1} - n \mathbb{T}^{n-2}\\
& = & \frac{\lef^n - 1}{\lef - 1 }- \frac{{(\lef -1)}^n - (- 1)^n}{\lef} - n {(\lef - 1)}^{n-2}.
\end{eqnarray*}
\end{examp}

There is an alternative characterization of $K(\quasiproj)$ due to Bittner, which is also sometimes 
useful in the context of Feynman graphs and motives.
\begin{dfn}[definition 2.3 of \cite{neeraja}]\label{Kvaralt}
$K^{\textup{bl}}(\quasiproj)$ is the abelian group generated by smooth complete varieties, subject to the conditions
\begin{enumerate}
\item{$[\emptyset] = 0$ and}
\item{$[\blow_Y X] = [X]-[Y] +[E]$ where $Y$ is a smooth and complete subvariety of $X$ and $\blow_Y X$ is the blowup of $X$ along $Y$ with the exceptional divisor of the blowup $E$.}
\end{enumerate}
\end{dfn}
We need to show that the definitions \ref{Kvar} and \ref{Kvaralt} do coincide, following \cite{neeraja}. To do this we need to show that there exists a map $K(\quasiproj) {\ra} K^{\textup{bl}}(\quasiproj)$ which is an isomorphism. This is constructed as an induced map from a map $\epsilon$ on varieties satisfying $e(X \setminus Y) = e(X) - e(Y)$. The map $e$ is constructed in the following way: Let $\overline{X}$ be the smooth completion of a smooth connected variety $X$ and $D= X \setminus \overline{X}$ a normal crossing divisor. Let us define the map $e(X) =\sum (-1)^l [D^{(l)}]_{\textup{bl}}$ with $D^{(l)}$ denoting the disjoint union of $l$-fold intersections of irreducible components of $D$. ($[-]_{\textup{bl}}$ denotes a class in $K^{\textup{bl}}(\quasiproj)$.) The fact that $e(X)$ is independent of the choice of completion $\overline{X}$ follows from the weak factorization theorem. The fact that the $e(X) = e(X \setminus Y) + e(Y)$ is proved by choosing a smooth and complete $X \subset \overline{X}$ such that $D = \overline{X} \setminus X$ is simple normal crossing and the closure $\overline{Y}$ in $X$ is smooth and normal-crossing with divisor $D$. ($D \cap Y$ is simple normal crossings divisor in $Y$.)
\par
Yet another useful result, which can be useful in applications in the context of Feynman graphs, is the following.
\begin{propn}[\cite{neeraja}]\label{kblowfor}
Let $X$ be a smooth connected variety, $Y \subset X$ a smooth connected subvariety of $X$ of codimension $d$. Let $E$ be the exceptional divisor of the blowup $\blow_Y X$ of $X$ along $Y$. Then
$$[\blow_Y X] = [X]+ \lef [E] - {\lef}^d [Y].$$
\end{propn}
A consequence of proposition \ref{kblowfor} is the notion of \emph{Tate twist}: there is a ring involution
\begin{equation*}
\lef \mapsto  \lef^{-1} = \textrm{ and } [X] \mapsto {\lef}^{- \dim X}[X] = \Q(1)^{\dim X} [X].
\end{equation*}
This is the true meaning of the Tate twist alluded above. In cohomological calculations, the product of $\Q(1)^{\dim X}$ and $[X]$ is replaced by a tensor product of a one dimensional rational vector space (``working mod torsion") raised to the dimension of the variety whose cohomology is being computed and a piece of the cohomology. In fact this correspondence between classes in $K(\quasiproj)$ and cohomology is one important 
aspect of the theory of motives. 
\par 
The reader may wonder what $K(\quasiproj)$ explicitly is and how does it depend on $k$. The fact of the matter is that this is a very hard question though certain things are known. For example, we have
\begin{thm} [Poonen, theorem 1 of \cite{poonen}] \label{poonen}
Let $k$ be a field of characteristic zero. Then $K(\quasiproj)$ is not an integral domain.
\end{thm}
In fact when $k = \mathbb{C}$, we have
\begin{thm} [Larsen--Lunts, theorem 2.3 of \cite{larsen}] \label{larsen}
Let $I$ be the ideal generated by the affine line, i.e., $I = \lef$ and denote by $SB$ the monoid of classes of stable birational complex varieties (the monoid structure coming from fiber products of varieties.) Then
$$K(\quasiproj) / I \simeq \Z[SB].$$
\end{thm}
(Two varieties $X$ and $Y$ are \emph{stably birational} if $X \times \mathbb{P}^m$ is birational to $Y \times \mathbb{P}^n$ for $m,n \geq 0$.)
\par
In Sahasrabudhe's thesis \cite{neeraja}, theorem \ref{larsen} is proved using Bittner's definition \ref{Kvaralt}.
\par
The main obstruction in obtaining results like theorems \ref{poonen} and \ref{larsen} in characteristic $p$ is that those results depend crucially on the resolution of singularities and the weak factorization of birational morphisms, statements that are not yet known to hold true in characteristic $p$.

\subsection{A Grothendieck ring for supermanifolds.}
One can generalize the construction of the Grothendieck ring of $\quasiproj$ in a quite straightforward way for the category of \emph{complex supermanifolds}. (See \cite{mansuper} for a standard introduction to supermanifolds.) 
\begin{dfn}[definition 2.2 of \cite{marrej}] \label{SGrV}

Let $\textbf{SVar}_\C$ be the category of complex supermanifolds. Let $K(\textbf{SVar}_\C)$ denote the free abelian group generated by the isomorphism classes of objects $\mathcal{X} \in \obj(\textbf{SVar}_\C)$ subject to the following relations. Let $F:\mathcal{Y} \hookrightarrow \mathcal{X}$ be a closed embedding of supermanifolds. Then 

\begin{equation*}\label{relSGr}
[\mathcal{X}]=[\mathcal{Y}]+[\mathcal{X} \setminus \mathcal{Y}],
\end{equation*}
where $\mathcal{X} \setminus \mathcal{Y}$ is the supermanifold 
\begin{equation*} 
\mathcal{X} \setminus \mathcal{Y} = (X \setminus Y, \mathcal{A}_X |_{X \setminus Y}).
\end{equation*}
\end{dfn}
The standard notation $\mathcal{A}_X |_{X \setminus Y}$ denotes restriction of the sheaf of supercommutative rings on $\mathcal{X}$ to the compliment. We can relate $K(\textbf{SVar}_\C)$ to $K(\textbf{Var}_\C)$ through
\begin{propn}[corollary 2.4 of \cite{marrej}] \label{ringSGr}
The Grothendieck ring $K(\textbf{SVar}_\C)$ of supervarieties is a
polynomial ring over the Grothendieck ring of ordinary varieties of
the form
\begin{equation*}\label{SGrPoly}
K(\textbf{SVar}_\C) = K(\textbf{V}_\C)[T],
\end{equation*}
where $T=[\mathbb{A}^{0|1}]$ is the class of the affine superspace of dimension $(0,1)$.
\end{propn}
The notion of birational and stable birational equivalence can be generalized also in a straightforward way to supermanifolds: two supermanifolds $\mathcal{X}$ and $\mathcal{Y}$ are said to stable birationally equivalent if there are superprojective spaces $\mathbb{P}^{n|m}$ and $\mathbb{P}^{r|s}$ such that $\mathcal{X} \times \mathbb{P}^{n|m}$ is birationally equivalent to $\mathcal{Y} \times \mathbb{P}^{r|s}$. Denote by $\Z[SSB]$ the monoidal ring of such supermanifolds. Then we have a result similar to theorem \ref{larsen}.
\begin{propn}[corollary 2.5 of \cite{marrej}] \label{SGrSSB} There is an isomorphism
\begin{equation*} 
K(\textbf{SVar}_\C) /I \simeq \Z[SSB],
\end{equation*}
where $I$ is the ideal generated by the classes $[\mathbb{A}^{1|0}]$ and
$[\mathbb{A}^{0|1}]$. 
\end{propn}
The main application of the Grothendieck rings constructed in this section is in defining a large class of topological and arithmetic invariants as measures on these rings. We take up this subject in section \ref{motzeta}.

\section{The Tannakian formalism} \label{tannaka}
\subsection{Categorical notions}
Most of the material in this section is based on \cite{delignetannaka} and Breen's survey of Saveedra-Rivano's thesis under Grothendieck \cite{sav}. 
In a nutshell, the main idea behind a Tannakian category is to equip an abelian category with a natural functor to vector spaces such that a fiber over each object in this category is a finite dimensional vector space.
\par
Let $(\Cat, \otimes)$ be a symmetric monoidal category.
\begin{dfn}[Internal homs]\label{inthom}
Let $X,Y \in \obj (\Cat)$ and consider the functor
\begin{eqnarray*}
F :\Cat & \ra & \sets,\\ 
T & \mapsto & \Hom(T\otimes X,Y).
\end{eqnarray*}
If $F = \Hom(-, K)$ where $K$ is some object in $\Cat$, then define $\underline{\Hom}(X,Y) =: K$. 
\end{dfn}
\begin{rem}\label{inthomrem}
Explicitly, $\Hom(T\otimes X, Y) = \Hom(T, \underline{\Hom}(X,Y))$. This spells out the fact that the functor $F$ is \emph{representable}.
\end{rem}
\begin{examp} Consider the category of all $R$-modules $\modu$. This is an obvious tensor category with $\point = R$. In $\modu$, $\underline{\Hom}(X,Y) = \Hom_{\modu}(X,Y)$. To see this, define the functor $F$ of definition \ref{inthom} to be $T \mapsto \Hom(T \otimes X,Y)$. Hom-tensor adjointness states $\Hom_{\modu}(T \otimes X,Y) = \Hom_{\modu}(T, \Hom_{\modu}(X,Y))$ and hence $F = \Hom_{\modu}(-, \Hom_{\modu}(X,Y))$ which, by definition, proves that $\underline{\Hom}(X,Y) = \Hom_{\modu}(X,Y)$.
\end{examp}
Let $\textup{ev}_{X,Y}: \underline{\Hom}(X,Y) \otimes X \ra Y$ be the morphism corresponding to $\id_{\underline{\Hom}(X,Y)}$. To see that this correspondence does make sense, take $T =\ihom(X,Y)$ in definition \ref{inthom} to get $$\Hom(\ihom(X,Y) \otimes X,Y) = \Hom(\ihom(X,Y), \ihom(X,Y)).$$
Define the dual of an object $X$ as $\hat{X} := \ihom (X, \point)$. For example in $\modu$ we have the following diagram
\begin{equation*}\begin{CD}
\ihom(X,Y) \otimes X @= \Hom(X,Y) \otimes X \\
@V{\textup{ev}_{X,Y}}VV           @VV{\textup{ev}_{X,Y}}V\\
Y @= Y
\end{CD} \end{equation*}
and $\textup{ev}_{X,Y}$ is given by $f \otimes x = f(x)$, the usual evaluation map in $\modu$. This shows that the abstract notion of evaluation makes sense.
\begin{dfn}[Reflexive objects] Suppose we are given the following digram
\begin{equation*}\begin{CD}
\ihom(X,Y) @= \ihom(X,Y)\\
@A{\psi}AA           @AA{\textup{ev}_{X,\point}}A\\
X\otimes \hat{X} @>{\textup{ev}_{X,\point} \circ \psi}>> \point
\end{CD} \end{equation*}
where $X$ is an object in some tensor category. If $\textup{ev}_{X,\point} \circ \psi$ is an isomorphism, we say that $X$ is reflexive. If all objects in a tensor category are reflexive, we call the category reflexive.
\begin{rem}
We note that the above definition does correspond to our usual notion of reflexiveness: from the definition of internal hom,
$$\Hom(X \otimes \hat{X}, \point) = \Hom(X, \ihom(\hat{X}, \point)).$$ If 
$\textup{ev}_{X,\point} \circ \psi$ is an isomorphism, we get a correspondence
\begin{equation*}
\textup{ev}_{X,\point} \circ \psi \longleftrightarrow X \stackrel{\sim}{\ra} \hat{\hat{X}}
\end{equation*}
with $\textup{ev}_{X,\point} \circ \psi \in \Hom(X \otimes \hat{X}, \point)$ and $X \stackrel{\sim}{\ra} \hat{\hat{X}}  \in \Hom(X, \ihom(\hat{X}, \point))$.
\end{rem}
Of course, not all categories are reflexive. For example, a one-line argument shows that $\modu$ is not. (Take $R = \Z$ and consider $\Z / 2\Z$.) 
\par
Putting all of these notions together, we have
\begin{dfn}[Rigid tensor category]\label{rigid}
A tensor category $\Cat$ is said to be \emph{rigid} if
\begin{enumerate}
\item{$\ihom(X,Y)$ exists for every $X,Y \in \obj (\Cat)$.}
\item{Functoriality of internal homs: All natural maps
$$\bigotimes_{i \in I} \ihom(X_i,Y_i) \lra \ihom \Big(\bigotimes_{i \in I} X_i, \bigotimes_{i \in I} Y_i \Big)$$ are isomorphisms.}
\item{$(\Cat, \otimes)$ is a reflexive category.}
\end{enumerate}
\end{dfn}
We need a last assumption for the main definition, namely that a tensor category be \emph{abelian}. We quickly review the definition: A category $\textbf{A}$ is \emph{abelian} \cite{gelfandmanin} is
\begin{itemize}
\item{$\Hom(X,Y)$ is an abelian group for all $x,y \in \obj(\textbf{A})$ and composition of morphisms is biadditive.}
\item{There exists a zero object $0$ such that $\Hom(0,0) = \emptyset$.}
\item{For all $X,Y \in \obj(\textbf{A})$, there exists an object $Z \in \obj(\textbf{A})$ and morphisms $\iota_X: X \ra Z$, $\iota_Y: Y \ra Z$, 
$\pi_X: Z \ra X$ and $\pi_Y: Z \ra Y$ with $\pi_X \circ \iota_X = id_X$, $\pi_Y \circ \iota_Y = id_Y$ and $\iota_X \circ \pi_X + \iota_Y \circ \pi_Y = id_Z$ and $\pi_X \circ \iota_Y = \pi_Y \circ \iota_X = 0$.}
\item{For any $f \in \Hom(X,Y)$ there is a sequence $K \stackrel{k}{\ra} X \stackrel{i}{\ra} I \stackrel{j}{\ra} Y \stackrel{c}{\ra}$ where $K$ (resp. $C$) is the kernel of $f$ (resp. cokernel of $f$) and such that $j \circ i = f$ and $I$ is both kernel of $c$ and cokernel of $k$.}
\end{itemize}
With this last piece at hand, we have
\begin{dfn}[Neutral Tannakian category] \label{ntc}
Let $k$ be a field of arbitrary characteristic. A neutral Tannakian category over $k$ is a abelian rigid tensor category $\Cat$ with a $k$-linear exact faithful functor called the \emph{fiber functor} $\omega: \Cat \ra \vect$ with $\vect$ the category of $k$-vector spaces\footnote{Some authors, for example Andre \cite{andre}, demand an additional $\Z/2\Z$-grading on $\vect$.}.
\end{dfn}
To fix notation: a neutral Tannakian category would be denoted as a tuple $(\Cat, \otimes, \omega)$ where $\omega(X \otimes Y) = \omega (X) \otimes \omega (Y)$ from the definition \ref{ntc}. Faithfulness, exactness and $k$-linearity are the usual notions.

\subsection{The structure of $\repG$} Let us come back to example \ref{repring} with $G$ an \emph{affine group scheme} instead, defined over $k$ a field of arbitrary characteristic. Let $V$ be a finite dimensional vector space over $k$. A representation of $G$ is a morphism of affine schemes $G \times V \ra V$, denoted as $\rho$. The category $\repG$ is defined in the following way.
\begin{itemize}
\item{Objects of $\repG$ are representations $(\rho, V)$.}
\item{Morphisms between two representations are intertwiners. Let $(\rho_1, V_1)$ and $(\rho_2, V_2)$ be two representations. $\Hom_{\repG}((\rho_1, V_1), (\rho_2, V_2))$ consists of maps $f: V_1 \ra V_2$ such that for every $g\in G$, the following diagram commutes:
\begin{equation*}\begin{CD}
G \times V_1 @>\rho_1>> V_1\\
@V{\id_{G} \times f}VV           @VV{f}V\\
G \times V_2 @>{\rho_2}>> V_2
\end{CD} \end{equation*}
}
\end{itemize}
The tensor structure is given by $\rho_1 \otimes \rho_2: G \ra \Aut(V_1 \otimes_k V_2)$, the unit object $\point: G \ra \Aut(\bf{1})$ where $\bf{1}$ is the one-dimensional vector space with trivial $G$-action. (A more general construction of $\repG$ over $R$, a commutative ring with identity, ivolves $V$ being a projective $R$-module of finite rank.)
\par
By a general well-known argument, $\repG$ is equivalent to the category of commutative (though not necessarily cocommutative) Hopf algebras\footnote{This is a fact of vital importance in the Connes-Kreimer and Connes-Marcolli theories of renormalization as a Riemann-Hilbert problem. }.

\par
The main theorem along the lines of the classical Pontryajin duality is
\begin{thm}[Tannaka-Krein theorem]\label{tannakakrein}
$\repG$ is a neutral Tannakian category with $\omega$ given by the forgetful functor. Furthermore $\TenAut (\omega) \simeq G$.
\end{thm}
The notation $\TenAut(\omega)$ denotes the natural transformations of the functor $\omega$ to itself preserving the tensor structure. 

\subsection{An equivalent definition} There is a definition of a neutral Tannakian category, due to Deligne \cite{delignetannaka}, which is equivalent to definition \ref{ntc}. This definition has the benefit of being more explicit. I present it for the benefit of the reader who may want to gain a shift in perspective.
\par
Let $(\Cat, \otimes)$ be a symmetric monoidal category over a field $k$ with a unit object $\point$. (The $k$-linearity of the functor $\otimes$ is understood.) Furthermore \emph{assume} $(\Cat, \otimes)$ to be abelian.

\begin{dfn}[Rigid tensor category, alternate]\label{rigidalt} The category $(\Cat, \otimes, \point)$ is \emph{rigid} if
\begin{enumerate}
\item{$\End(\point) = \Hom(\point, \point) \simeq k$,}
\item{For all $X \in \obj(\Cat)$ there exist objects $\hat{X} \in \obj(\Cat)$ and morphisms $\delta: \point \ra \hat{X} \otimes X$ and $\textup{ev}: X \otimes \hat{X} \ra \point$ such that
\begin{eqnarray*}
X \stackrel{\id \otimes \delta}{\lra} X \otimes \hat{X} \otimes X \stackrel{{\textup{ev}} \otimes \id}{\lra} X,\\
\hat{X} \stackrel{\delta \otimes \id}{\lra} \hat{X} \otimes X \otimes \hat{X} \stackrel{\id \otimes {\textup{ev}}}{\lra} \hat{X}.
\end{eqnarray*}}
\end{enumerate}
\end{dfn}
We now \emph{define} the internal hom to be $\ihom(X,Y):= \hat{X} \otimes Y$ for all $X,Y \in \obj(\Cat)$. One can verify, upon making the identification $\Hom(X, Y) = \Hom(X, Y \otimes \point)$ that this coincide with definition \ref{inthom}. The dual is a functor $\Cat \to \Cat$ given by $\hat{(-)} = \ihom(-,\point)$. It exists if and only if $\ihom(X, -)$ exists and $\ihom(X,\point) \otimes Y \stackrel{\sim}{\lra} \ihom(X,Y)$.
\begin{table}
\[
\begin{tabular}{|l|l|l|}
\hline
Category & Unit & Dual\\
\hline
$\vect$ & $\textbf{1}$ & dual vector space $\hat{V}$\\
$\repG$ for $G$ affine group scheme & $k$ & $\hat{V}$ with induced $G$-action\\
Flat $\C$-vector bundles on $X$ & ${\underline{\C}}_X$ & $\widehat{\mathcal{E}} = \Hom(\mathcal{E},{\underline{\C}}_X)$ \\
Connections on $\mathbb{G}_m$ & $\mathbb{G}_a$ & $(\widehat{\mathcal{E}}, \widehat{\nabla})$\\
\hline
\end{tabular}
\]
\caption{Examples of rigid tensor categories}\label{rigidTCexamp}
\end{table}
Deligne's definition of a neutral Tannakian category is the same as definition \ref{ntc}, namely, an abelian rigid tensor category with a fiber functor $\omega: \Cat \lra \vect$, the functor being exact, faithful, $k$-linear and preserving the tensor structure.
\par
The main theorem, in any case, is a stronger formulation of theorem \ref{tannakakrein}, variously attributed to Deligne, Grothendieck and Saavedra-Rivano:
\begin{thm}\label{tannakastrong}
Let $\textbf{T}$ be a neutral Tannakian category over a field $k$. Then
$\TenAut(\omega)$ is an affine group scheme over $k$ and we have the equivalence of categories
$$\textbf{T} \simeq {\textbf{Rep}}_k \Big(\TenAut(\omega)\Big),$$
where ${\textbf{Rep}}_k (-)$ is the category of $k$-linear representations. Furthermore, if ${\textbf{T}} \simeq \repG$ for some affine group scheme $G$, then there exists an exact, faithful, $k$-linear tensor functor $\omega$ such that $$G \stackrel{\sim}{\lra} \TenAut(\omega).$$
\end{thm}
The functor $\omega$ is representable, following the remark \ref{inthomrem}. Ultimately, the rigidity features guarantee the group structure. In fact, the definition of $\omega$ in terms of ``functor-of-points" suggests we look at $\omega$ with $\sets$ as the target category; this provides an interpretation of the action of the group $\TenAut(\omega)$ as a fundamental group, a theme we take up next. 

\subsection{Fundamental groups of schemes} Tannakian categories can also be used  as a language to understand a beautiful theory developed by Grothendieck, Artin and others, of the relationship between fundamental groups of \'etale coverings of a scheme and Galois groups. The canonical reference for this material is SGA I \cite{SGA1}. 
\par
\begin{dfn}[\'Etale covering of a scheme]\label{etalecover}
Let $Y$ be a scheme over $k$. An \'etale covering $Y \ra X$ is an affine morphism given by $A \ra B$ with $X= \Spec A$ and $Y =\Spec B$, $A$ and $B$ algebras, and such that
\begin{enumerate}
\item{$B$ is flat over $A$,}
\item{${\textup{Der}}_{A} B = 0$ and}
\item{$B$ is finite over $A$.}
\end{enumerate}
\end{dfn}
\begin{examp}
Let $X$ be a smooth variety over $\C$. Saying that $Y \ra X$ is an \'etale covering over $X$ is the same as $Y_\C \ra X_\C$ is a covering in the usual sense.
\end{examp}
\begin{examp}
Let $X = \Spec k$. Then we have the equivalence of sets (in fact of categories!):
\begin{equation*}
\textrm{\{connected \'etale coverings of X\} } \simeq \textrm{ \{finite seperable field extensions of $k$\}}.
\end{equation*}
\end{examp}
For a given and fixed scheme $X$, \'etale coverings form a category with morphisms $Y \ra Y'$ and such that the following diagram
\begin{equation*}\begin{CD}
Y @>>> Y'\\
@VVV           @VVV\\
X @= X
\end{CD} \end{equation*}
commutes for \'etale coverings $Y \ra X$ and $Y' \ra X$. Denote this category as ${\etalecov}_X$. Given a geometric point $\bar{x} \in X$, define a functor
\begin{eqnarray*}
\omega: {\etalecov}_X & \lra & \sets,\\
(Y \stackrel{\pi}{\ra} X) & \mapsto & \pi^{-1}(\bar{x}).
\end{eqnarray*}
\begin{rem}
In case of the universal covering $\tilde{X} \ra X$ has the property that $\Hom(\tilde{X}, X) = \omega(Y)$ does not exist but 
$$ \omega(Y) = \limproj_i \Hom(X_i, Y),$$
This motivates our next definition.
\end{rem}
\begin{dfn}[Fundamental group]\label{fundgp}
\begin{equation*}
\pi_1(X, \bar{x}) := \limproj_i \Aut_X (X_i).
\end{equation*}
\end{dfn}
\begin{examp}Let $X$ be a complex variety. Then $\pi_1(X) = \widehat{\pi_1(X_{\C}, x_{\C})}$.
\end{examp}
More interestingly,
\begin{examp}
Let $X = \Spec k$. Then $\pi_1(X) = \Gal (\bar{k}/k)=: G_k$.
\end{examp}
In fact, we have, following definition \ref{fundgp}:
\begin{thm}[Grothendieck]\label{finsetsetale} The category of finite \'etale schemes over $k$ is equivalent to the category of finite sets with continuous $G_k$-action.
\end{thm}
In fact, in a Tannakian category with the fiber functor $\omega$, the theorem \ref{finsetsetale} can be understood as saying that the fundamental group acts as automorphisms of $\omega$. To see this, view the category of flat vector bundles over a scheme $X$ as being equivalent to the category of representations of $\pi_1(X,x)$. Furthermore, the category is recovered by studying the category of representations of $\pi_1(X,x)$ using theorem \ref{tannakastrong}.
\par
I remark that theorem \ref{finsetsetale} is a special case of the general \emph{Grothendieck-Galois correspondence} which states that the category of finite \'etale $k$-schemes is equivalent to the category of finite sets with continuous $G_k$-action. 
\subsection{The function-sheaf correspondence} The function-sheaf correspondence is a profound application of constructibility to obtain all ``interesting" functions on a space over an arbitrary base ring in terms of certain sheaves on it. As such, it also connects with the \emph{Geometric Langlands program}. In this subsection I scratch the surface, closely following the notes of Sug Woo Shin \cite{shin}.
\par
Fix a prime $l$ and let $K$ be a finite extension of $\Q_l$. Denote by $\mathcal{O}_K$ the ring of integers of $K$. Let $X$ be a connected scheme over $k$ with $\textup{char }k \neq l$. Let $X_{et}$ be an \'etale site of $X$. Call the \'etale sheaf $\mathcal{G}$ \emph{locally constant} is $f|_{U}$ is locally constant for an \'etale covering $U$ of $X$. The sheaf $G$ is said to be \emph{constructible} is $X$ is constructible (i.e. can be written as disjoint union of locally closed subschemes of $X$) and $\mathcal{G}$ is such that it defines a locally constant sheaf, finite on each strata of $X$.
\begin{dfn}[locally constant $l$-adic sheaf]
An $l$-adic sheaf on $X_{et}$ is a projective system $(\mathcal{F}_n)_{n \in \N}$ of constructible sheaves $\mathcal{F}_n$ such that $\mathcal{F}_{n+1} \ra \mathcal{F}_n$ induces an isomorphism $\mathcal{F}_{n+1} \otimes (\Z / l^n) \simeq \mathcal{F}_n$. An $l$-adic sheaf is \emph{locally constant} if each $\mathcal{F}_n$ is so. 
\end{dfn} 
One defines a \emph{category of $K$-sheaves} as a category whose objects are constructible $\mathcal{O}_K$-sheaves (= constructible sheaves of $\mathcal{O}_K$-modules) and with
$$\Hom_{K-\textup{sheaves}}(\mathcal{F} \otimes K, \mathcal{G} \otimes K) := \Hom_{\mathcal{O}_K-\textup{sheaves}} (\mathcal{F}, \mathcal{G}) \otimes_{\mathcal{O}_K} K.$$
By taking a limit over $l$ of the categories of $K$-sheaves, we obtain the category of $\overline{\Q}_l$-sheaves. Locally constant $\overline{\Q}_l$-sheaves are limits of locally constant $\mathcal{O}_K$-sheaves.
\begin{dfn} [$l$-adic local system]
A locally constant $\overline{\Q}_l$-sheaf is called an $l$-adic local system.
\end{dfn} 
Let $\overline{x} \in X$ be a geometric point. Let $\mathcal{E}$ be an $l$-adic local system on $X$. The stalk of $\mathcal{E}$ is defined as the direct limit of stalks of $K$-sheaves. These are, in turn, defined in the following way: for $\mathcal{F}$ a $\mathcal{O}_K$-sheaf on $X$, the stalk at $\overline{x}$ is $\limproj_n (\mathcal{F}_n)_{\overline{x}}$ where $(\mathcal{F}_n)_{\overline{x}}$ are the usual stalks on the \'etale site. Denote the stalk of $\mathcal{E}$ at $\overline{x}$ as $\mathcal{E}_{\overline{x}}$. We say that the local system $\mathcal{E}$ has \emph{rank $r$} if $\mathcal{E}_{\overline{x}}$ contains $r$-copies of $\overline{\Q}_l$. 
\begin{thm}
Let $\overline{x} \in X$ be a geometric point of a finite connected scheme $X$. The functor $\mathcal{F} \ra \mathcal{F}_{\overline{x}}$ makes the category of $l$-adic local systems on $X$ equivalent to the category of continuous representations of $\pi_1(X, \overline{x})$ on finite dimensional $\overline{\Q}_l$ vector spaces.
\end{thm}
Compare this with theorem \ref{finsetsetale}: in both cases, action of the automorphisms of the fiber functor are interpreted in terms of ``nice" (= \'etale!) geometric categories. 
\par
We end this discussion by noting a special case of the function-sheaf correspondence: let $H$ be a (connected, seperated) commutative group scheme of finite type over $\F_{p^n}$ with $p \neq l$ and let $\mathcal{E}$ be an $l$-adic local system on $H$. Let $m^{*}$ denote the pull-back of the multiplication map $m: H \times_{\F_{p^n}} H \ra H$. We call $\mathcal{E}$ an $l$-adic \emph{character sheaf} if 
$m^{*} (\mathcal{E}) \simeq p^{*}_1(\mathcal{E}) \otimes p^{*}_2(\mathcal{E})$. Here $p_1$ and $p_2$ are the projection morphisms $H \times_{\F_q} H \ra H$.
\begin{thm}[Function-sheaf correspondence]\label{functionsheaf}
We have a natural bijection of sets
$$ \Hom_\abgps(H(\F_q), \overline{\Q}_l) \leftrightarrow \{l\textrm{-adic character sheaves on }H\} .$$
\end{thm}
\begin{rem}
A basic fact: any commutative group scheme over a field $k$ is always an extension of an abelian variety by an affine group $k$-scheme (= a commutative $k$-Hopf algebra.) Therefore, we loose nothing by following the notation and thinking of $H$ as an abelian variety. For example, the reader may think of $H$ as the Picard group of a curve, to fix ideas.
\end{rem}
The proof of theorem \ref{functionsheaf} uses the functoriality of the action of the absolute Frobenius\footnote{Let $X$ be a (connected) scheme over $\F_{p^n}$. By the ``absolute Frobenius" we mean an $\F_{p^n}$-morphism which is the identity on $X$ as a topological space and is the map $x \mapsto x^{p}$ on the structure sheaf ${\mathcal{O}}_X$.} on the stalks of $l$-adic local system $\mathcal{E}$ (cf. \cite{shin} for a nice sketch of the proof.)
\par
A speculation: Upon a fixed identification $\overline{\Q}_l \stackrel{\sim}{\ra} \C$, one way to think of theorem \ref{functionsheaf} is\footnote{There are several subtleties involving base change and defining the absolute Frobenius, so the reader should take this statement with a grain of salt.}: all character maps $\phi$ of a given commutative Hopf algebra $\textbf{H}$ arise as appropriate $l$-adic character sheaves on $\textbf{H}$. It remains a very interesting project to actually implement this philosophy when $\textbf{H}$ is the Connes-Kreimer Hopf algebra of Feynman graphs\footnote{In the Connes-Kreimer theory, the coproduct of $\textbf{H}$ gives a recursive formula for the factorization of loops in the prounipotent complex Lie group $G(\C):= \Hom({\textbf{H}}, \C)$. The Birkhoff factorization of $\phi$ as algebra homomorphisms, needed for the counterterms, satisfy the Rota-Baxter identity.}  (\cite{ck1} \cite{ck2}) and $\phi$ a Feynman rule that assigns a generally divergent projective integral to a given Feynman graph and understanding its relationship with the Arapura motivic sheaves and renormalization as outlined in Marcolli's paper \cite{matilde}.

\section{Weil cohomology}
This section presents the notion of a Weil cohomology, a set of functorial properties any good cohomology theory of smooth projective varieties should satisfy. The section very closely follows the first parts of Kleiman's article in \cite{bible}.
\subsection{Main definition}\label{weil}
Let $k$ (resp. $K$) be fields of arbitrary characteristic (resp. zero characteristic.) Let $\smproj$ be the category of smooth projective schemes over $k$ with smooth scheme maps as morphisms. Let $\grvect$ be the category of $\Z$-graded anticommutative $K$-algebras with $K$-algebra maps that preserve gradings as morphisms.
\par
Suppose $V \in \obj(\grvect)$. Therefore $V:= \bigoplus_{i \in \Z} V_i$. Set $V_0 = K$. A Weil cohomology is a contravariant functor
$$H^{*}: \smproj \lra \grvect$$
with $H^{*}(X)= \bigoplus_{n\in \Z} H^{i}(X)$ that satisfies the following properties:
\subsubsection{Finiteness} Each $H^{i}(X)$ has finite dimension and $H^{i}(X) = 0$ unless $0 \leq i \leq 2\dim X$.
\subsubsection{Poincar\'e duality} There are two equivalent versions. The first version goes as follows: Let $r:= \dim X$. For each such $X$, there is an isomorphism
$$H^{2r}(X) \stackrel{\sim}{\lra} K$$
and a nondegenerate pairing 
$$H^{i}(X) \times H^{2r-i}(X) \stackrel{\sim}{\lra} K.$$
The second version of the Poincar\'e duality goes as follows:
$$\widehat{H^{i}(X)} \simeq H^{2r-i}(X),$$
where $\widehat{H^{i}(X)} = \Hom(H^{i}(X), K)$.
\subsubsection{K\"unneth formula} Let $X \times_k Y$ be the fiber product of two objects $X$ and $Y$ of $\smproj$. Consider the following diagram of projections
\begin{equation*}\begin{CD}
X \times_k Y @>{\pi_X}>> X \\
@V{\pi_Y}VV          \\
Y 
\end{CD} \end{equation*}
The projections induce the isomorphism
$$H^{*}(X) \otimes_K H^{*}(Y) \simeq H^{*}(X \otimes Y).$$
NB: since we are working over fields, the usual ``$\Tor$" term is absent.
\subsubsection{Cycle maps} Let $C^i(X)$ denote the free abelian group generated by closed irreducible subschemes of $X$ of codimension $i$. In a Weil cohomology, there should be a group homomorphism
$$\gamma^{i}_{X}: C^{i}(X) \lra H^{2i}(X)$$
satisfying the following properties:
\begin{enumerate}
\item{[Functoriality] There are two functorial conditions on cycle maps. The first (``pull back") goes as follows: Let $r := \dim X$. Let $f: X \ra Y$ be a morphism of $\smproj$. Let $C^{i}(X)$ and $C^{i}(Y)$ denote the group of cycles of codimension $i$ on $X$ and $Y$ respectively. The following diagram commutes:
\begin{equation*}\begin{CD}
C^{i}(Y) @>{f^{*}}>> C^{i}(X) \\
@V{\gamma^i_Y}VV           @VV{\gamma^i_X}V\\
H^{2i}(Y) @>{f^{*}}>> H^{2i}(X)
\end{CD} \end{equation*}
(The top map is a pull-back on cycles.)
\par
The second functorial condition (``push forward") goes as follows: Let $s:= \dim Y$. Now consider the pull back map defined above: if $$H^{2i}(Y) \stackrel{f^*}{\lra} H^{2i}(X),$$ then by Poincar\'e duality, we obtain the push-forward map 
$$H^{2r-2i}(X) \stackrel{f_{*}}{\lra} H^{2s- 2i}(Y)$$
such that the following diagram commutes:
\begin{equation*}\begin{CD}
C^{r-i}(X) @>{f_{*}}>> C^{s-i}(X) \\
@V{\gamma^{r-i}_X}VV           @VV{\gamma^{s-i}_Y}V\\
H^{2r-2i}(X) @>{f_{*}}>> H^{2s-2i}(Y)
\end{CD} \end{equation*}
(The map on the top denotes push-forward on cycles.)}
\item{[Multiplicativity] The map $\gamma^{i+j}_{X \times_k Y} = \gamma^{i}_{X}(Z) \otimes_K \gamma^{j}_{Y}(W)$.}
We see, by applying the definition of the cycle map and the K\"unneth formula that 
\begin{eqnarray*}
Z \times_k W \in C^{i+j}(X \times_k Y) &\stackrel{\gamma^{i+j}_{X \times_k Y}}{\lra}& H^{2(i+j)}(X \times_k Y)\\
&\stackrel{\sim}{\lra}& \bigoplus_i H^{2(i+j)-l} H^{2(i+j)-l}(X) \otimes_K H^{l}(Y)
\end{eqnarray*}
so the multiplicativity axiom makes sense.
\item{[Calibration] Let $P$ be a point. Then the cycle map $\gamma_P: C^0 \ra H^0(P)$ is the same as the inclusion of $\Z$ in $K$.} 
\end{enumerate}
\subsubsection{Lefschetz theorems} There are two versions of the Lefschetz theorems, one ``weak" and the other ``strong". Let $h: W \ra X$ be the inclusion of a smooth hyperplane section $W$ of some smooth projective scheme $X$ of dimension $r$.
\begin{enumerate}
\item{[Weak Lefschetz] The induced map $h^{*}: H^{i}(X) \ra H^{i}(W)$ is an isomorphism for $i \leq r -2$ and an injection for $i = r-1$.}
\item{[Hard Lefschetz] Define the \emph{Lefschetz operator} 
\begin{eqnarray*}
L: H^{i}(X) & \ra & H^{i+2}(X)\\
	Lx  & \mapsto & x \cdot \gamma^{1}_X(W).
\end{eqnarray*}
Then for $i \leq r$, $L^{r-i}: H^{i}(X) \lra H^{2r-i}(X)$ is an isomorphism.}
\end{enumerate}
In summary: a Weil cohomology theory over $k$ is a contravariant functor from the category $\smproj$ to the category $\grvect$ with a nondegenerate pairing  satisfying a Poincar\'e duality, with a K\"unneth formula, with cycle maps that are functorial with respect to pull-backs and push-forwards and satisfying the Lefschetz theorems.

\subsection{Some properties of cycle maps} There are two important properties of cycle maps in the context of pure motives.
\subsubsection{Intersection pairing and cup products} Let $Z$ and $W$ be two properly intersecting cycles in $X$ and let $Z \cap W$ denote the intersection pairing. Let $\Delta: X \ra X \times X$ be the diagonal map. Then $\gamma_X(Z \cap W) = \gamma_{X \times X} \circ \Delta^{*}(Z \times W)$ where $\Delta^{*}$ denotes the pull-back of $Z \times W$. We knew from the functoriality of cycle maps that this equals $\Delta^{*} \circ \gamma_{X \times X} (Z \times W) = \Delta^{*}(\gamma_X(Z) \otimes \gamma_X(W))$ by the multiplicativity of the cycle maps. So we get
$$\Delta^{*}(\gamma_X(Z) \otimes \gamma_X(W)) = \gamma_X(Z) \cdot \gamma_X(W)$$
by applying pull-back again.
\subsubsection{Correspondences as operators} Let $r := \dim X$. Observe from the definitions that
\begin{eqnarray*}
H^{i}(X \times Y) &=& H^{i}(X) \otimes H^{i}(Y)\\
		  &=& \widehat{H^{2r-i}(X)} \otimes H^{i}(Y)\\
		  &=& \Hom(H^{2r-i}(X), K) \otimes H^{i}(Y)\\
		  &=& \Hom(H^{2r-i}(X), H^{i}(Y)).
\end{eqnarray*}
Therefore, we can view each element of $H^{i}(X \times Y)$ as an \emph{operator} from $H^{2r-i}(X)$ to $H^{i}(Y)$. Such operators are called \emph{correspondences} for the obvious reasons of section \ref{puremot}.

\subsection{The Standard Conjectures}\label{standconj} 
The original formulation is in Grothendieck's Tata lecture \cite{standconj}. I follow Murre's lecture \cite{murre}.
\par
Rescalling the indices in the hard Lefschetz theorem, the isomorphism can be equivalently written as $L^{r-i}: H^{r-j}(X) \ra H^{r+j}$ for $r = \dim X$. Using the map $L$ we can define a unique linear operator $\Lambda$ which makes the following: For $0 \leq j \leq r-2$ the diagram
\begin{equation*} 
\begin{CD}
H^{r-j}(X) @>{L^{j}}>> H^{r+j}(X) \\
@V{\Lambda}VV           @VV{L}V\\
H^{r-j-2}(X) @>{L^{j+2}}>> H^{r+j+2}(X)
\end{CD} 
\end{equation*}
commutes, and for $2 \leq j \leq r$
\begin{equation*} 
\begin{CD}
H^{r-j+2}(X) @>{L^{j-2}}>> H^{r+j-2}(X) \\
@A{L}AA           @AA{\Lambda}A\\
H^{r-j}(X) @>{L^{j}}>> H^{r+j+2}(X)
\end{CD} 
\end{equation*}
the diagram commutes.
We have
\begin{conj}[B(X)]\label{algebraic} The operator $\Lambda$ is algebraic. That is $$\Lambda(Z) = \gamma^{*}_{X \times X}(Z) \textrm{ for all }Z \in \ichow(X \times X) \otimes \Q.$$
\end{conj}
By the hard Lefschetz theorem, we have the isomorphisms $L^{r-i}: H^{i}(X) \ra H^{2r-i}(X)$. Let
$$P^{i}(X) := \ker\{L^{r-i-1}: H^{i}(X) \ra H^{2r-i+2}(X)\}$$
be the set of \emph{primitive elements} of $H^{i}(X)$. Let $x,y \in C^{i}(X) \cap P^{2i}(X)$ for $i \leq r/2$. 
\begin{conj}[Hdg(X)]\label{hodge}
Consider the pairing
$$(x,y) \mapsto (-1)^i \langle L^{r-2i}(x), y \rangle$$
where $\langle -,- \rangle$ denotes the cup product. This pairing is positive definite.
\end{conj}
It is known that conjecture \ref{hodge} is true for \'etale cohomology in characteristic zero. (The fact that \'etale cohomology is a Weil cohomology and that, in particular, the hard and weak Lefschetz theorems hold, requires some work to show.) 
\par
Conjectures \ref{algebraic} and \ref{hodge} taken together imply the Weil conjectures, among other things. One cares about the standard conjectures because they capture something intrinsic about the cycles independent of the Weil cohomology that the cycle map maps to. For example, we have the following, due to Grothendieck:
\begin{thm}\label{stimpl}
Assume conjectures \ref{algebraic} and \ref{hodge}. Then the Betti numbers $\dim H^{i}(X)$ is independent on the choice of $X$.
\end{thm}

\section{Classical motives} \label{puremot}
As noted in the introduction, one of the central goal of theory of motives is \emph{linearization} of the category of algebraic varieties (or schemes)  over an arbitrary base field (or ring). In fact, what we are after is obtaining something stronger than additivity (which is the ``coarsest" form of linearization)-- we want an abelianization of the category of algebraic varieties! One way to abelianize the category of varieties is to replace morphisms in this category by \emph{correspondences}.
\par
\subsection{Correspondences of curves} By a curve, we mean a nonsingular variety of dimension 1.
Let us start with the case of correspondences of curves over the complex numbers (classical theory due to Castelnuovo and Severi).
\begin{dfn}[section 2.5 of \cite{grifharr}] \label{corrcurve}
Let $C$ and $C'$ be two curves. A \emph{correspondence} of degree $d$ between $C$ and $C'$ is a holomorphic map 
\begin{equation*}
T: C \ra C', 
p \mapsto T(p)
\end{equation*} 
where $T(p)$ a divisor of degree $d$ on $C$. 
Furthermore, the \emph{curve of correspondence} between $C$ and $C'$ is given by the curve 
\begin{equation*}
D = \{(p,q)| q \in T(p) \} \subset C \times C'. 
\end{equation*}
Let $D \subset C \times C'$ be a curve. Then the correspondence associated to $D$ is defined by 
\begin{equation*}
T(p) = \pi^{*}_p(D) \in \textup{Div}(C')
\end{equation*}
where $\pi^{*}_p: C' \ra C \times C'$ is given by $q \mapsto (p, q)$. 
\end{dfn}
\begin{rem} The ``converse" part of definition (\ref{corrcurve}) states that we should view correspondences as formal linear combinations of all $0$-dimensional subvarieties on $C'$.
\end{rem}

\subsection{Equivalence relations on algebraic cycles}
All through we will work with smooth projective varieties over a field $k$. We denote the category of such objects as $\smproj$. Furthermore, most of the time, we'd restrict ourselves to such varieties that are also irreducible. Much of the material in this section (and the next) is from the excellent reviews by Murre \cite{murre} and Scholl \cite{schorev}.
Recall
\begin{dfn}
An \emph{algebraic cycle} $Z$ on $X$ of codimension $i$ is a formal linear combination of closed irreducible subvarieties $Z_\alpha$ of codimension $i$; thus,
\begin{equation*}
Z = \sum_\alpha n_\alpha Z_\alpha, n_\alpha \in \Z. 
\end{equation*}
\end{dfn}
Algebraic cycles (or cycles, for short) of a given codimension $i$ in a given
smooth projective $X$ form an abelian group with repect to formal sums. Denote such a group as $\mathcal{Z}^i(X)$. As it stands, $\mathcal{Z}^i(X)$ is very ``big", so we want to obtain a smaller group by taking a quotient of $\mathcal{Z}^i(X)$ by equivalence relations on cycles.
\par
There are several equivalence relations that one can impose on cycles, but we'd restrict ourselves to the following.
\begin{enumerate}
\item \label{rat}{(Rational equivalence) Let $Z$ be a cycle of a fixed codimension $i$ on $X$ irreducible of dimension $d$. We say $Z$ is rationally equivalent to zero and write $Z \rat 0$ if there exists $(Y_\alpha, f_\alpha)$ with $Y_\alpha$ irreducible of codimension $i-1$ and $f_\alpha$ a rational function on $Y_\alpha$ such that $\sum_\alpha v_{Y_\alpha} (f_\alpha) Y_\alpha = Z$. (Here $v$ denotes the valuation of the rational function.)}
\item \label{hom}{(Homological equivalence) Once again,  let $Z$ be a cycle of codimension $i$. Let $H^{*}_{B}(X, \Q)$ be the Betti cohomology of $X$. Then there exists a map (in fact a group homomorphism!) $Z \mapsto [Z]$ where $[Z]$ is a cohomology class in $H^{2i}_B (X, \Q)$. We say that $Z$ is homologically equivalent to zero and write $Z \homo 0$ if the image of $Z$ under this map vanishes. Furthermore, cycle maps are general gadgets that associate $i$-cycles to $2i$-cohomology classes in any good cohomology theory (not just Betti) and statisfying certain functorialilty conditions, a notion formalized by Weil cohomology, see \ref{weil}.  Homological equivalence is a notion that is to hold true in \emph{any} (and perhaps more strongly in \emph{all}) of these good cohomology theories.}
\item \label{num}{(Numerical equivalence) Let $Z$ as in above. We say $Z$ is numerically equivalent to zero and write $Z \num 0$ if the intersection number $\#(Z \cap Z') = 0$ for all $Z'$ cycles of the same codimension as $Z$. (To spell things out a bit more: the definition says that $\# (Z_\alpha \cap Z'_\beta) = 0$ for all $\alpha, \beta$ and where $Z = \sum_\alpha n_\alpha Z_\alpha$ and $Z' = \sum_\beta n_\beta Z'_\beta$.)}
\end{enumerate} 
It can seen that the relations given above are indeed equivalence relations. From now on, we would write $\sim$ to mean any of three relations above unless we fix the equivalence relation, in which case we would specify it as such.
\begin{dfn} Fix a smooth projective variety $X$.
The group $$C^{i}_\sim (X) := \mathcal{Z}^i (X) / \mathcal{Z}^i_{\sim} (X)$$
where $\mathcal{Z}^i_{\sim} (X)$ denotes the subgroup of cycles identified under the equivalence relation $\sim$.  When $X$ is irreducible and of dimension $d$, $C_\sim(X) := \oplus_{i =0}^{d} \cycle (X)$. When $\sim = \homo$, $C^{i}_\sim(X)$ is called the \emph{Chow group} and is denoted as $CH^i(X)$. Notation: we write $C_\sim (X)_\Q$ to mean $C_\sim(X) \otimes \Q$.
\end{dfn}
\begin{rem} Notice that (\ref{rat}) is just a generalization of the notion of divisors on curves and that of linear equivalence. The relation (\ref{num}) is tricker because the interesction pairing $Z \cap Z'$ may not be always defined. In this case, one resorts to various \emph{moving lemmas}.  Perhaps the most problematic is the equivalence relation (\ref{hom}) since the definition is based on a choice of a cohomology theory; however assuming one of the implications of the Standard conjectures along with Jannsen's result help solving this problem (see theorem \ref{jannsen}.)
\end{rem}
An important observation is 
$$C^i_{\rat}(X) \subseteq C^i_{\homo}(X) \subseteq C^i_{\num}(X) \subset \cycle(X)$$
going from the finest to the coarsest equivalence relations on cycles. An extremely important conjecture in the theory of classical motives is
\begin{conj}[Fundamental Conjecture D(X)]\label{fundamental}
$$C^i_{\homo}(X) = C^i_{\num}(X).$$
\end{conj}
The conjecture \ref{fundamental} follows from the Standard Conjectures of Grothendieck B(X) (conjecture \ref{algebraic}) and Hdg(X) (conjecture \ref{hodge}).

\subsection{Pure motives}
Just like in the case of curves (see definition (\ref{corrcurve}), we view correspondences on a smooth projective variety as formal linear combinations of closed irreducible subvarieties, i.e, as elements in \emph{group of correspondences} $C_{\sim} (X \times Y)_\Q =: \cor_{\sim} (X,Y)$.
\par
Let $f \in \cor_{\sim} (X,Y)$ and $g \in \cor_{\sim} (Y,Z)$. The composition of correspondences is given by 
\begin{equation*}
g \bullet f = \textup{pr}_{XZ} ((f \times Z) \cdot (X \times g)).
\end{equation*}
A emph{projector} $p \in \cor^{0}_{\sim}(X,X)$ is an idempotent $p \bullet p = p$.
\begin{dfn}[the category of effective pure motives]\label{effmot}
Fix an equivalence relation on cycles $\sim$. The category of effective pure motives over a field $k$ is denoted as $\mot^{+}_\sim(k)$ has
\begin{itemize}
\item{Objects: $(X,p)$ where $X \in \obj(\smproj)$ and $p$ a projector.}
\item{Morphisms: Let $M= (X,p)$ and $N= (Y,q)$. Then $\Hom_{\mot^{+}_\sim(k)}(M,N) := q \bullet \cor^{0}_{\sim}(X,Y) \bullet p$.}
\end{itemize}
\end{dfn}
Elements of $\Hom_{\mot^{+}_\sim(k)}(M,N)$ are simply the composition of correspondences $X \ra X \ra Y \ra Y$. By modifying the definition \ref{effmot} to allow for Tate twists, we get the category of virtual pure motives over $k$.
\begin{dfn}[the category of virtual pure motives]\label{virmot}
The category of virtual pure motives over a field $k$ is denoted as $\mot_\sim(k)$ has
\begin{itemize}
\item{Objects: $(X,p,m)$ where $X$ and $p$ as above and $m= \dim X$.}
\item{Morphisms: Let $M= (X,p,m)$ and $N= (Y,q, n)$ (and $n= \dim Y$). Then $\Hom_{\mot_\sim(k)}(M,N) := q \bullet \cor^{n-m}_{\sim}(X,Y) \bullet p$.}
\end{itemize}
\end{dfn}
Some trivial and distinguised pure motives: ${\bf 0}:= (\Spec k, \id, 0)$, the motive of a point, ${\bf L}:= (\Spec k, \id, -1)$ the Lefschetz motive and ${\bf T}:= (\Spec k, \id, -1)$ the Tate motive. 
\subsection{The motives functor} 
Recall that if $\psi: Y \ra X$ is a scheme morphism, then the \emph{graph} of $\psi$, $\Gamma_\psi := (\psi \times id_Y) \circ \Delta_Y$ where $\id_Y$ and $\Delta_Y$ are respectively the identity and diagonal maps on $Y$. That is, the graph is simply the composition of maps in the sequence
$$Y \stackrel{\Delta_Y}{\lra} Y \times Y \stackrel{\psi \times id_Y}{\lra} X \times Y.$$ The transpose $\Gamma^{t}_\psi$ is obtained by exchanging the factors of $X \times Y$.
\par
\begin{dfn}
The functor $$\mathfrak{m}_\sim: \smproj^{op} \ra \mot_\sim(k)$$ is defined in the following way: $\mathfrak{m}_{\sim}(X) := (X, \Delta_X,0)$ and $\mathfrak{m}_{\sim}(\psi) = \Gamma^t_{\psi}: \mathfrak{m}_{\sim}(Y) \ra \mathfrak{m}_{\sim}(X)$ and where $\Gamma^t_{\phi}$ is the transpose of the graph of $\psi: X \ra Y$ a morphism of smooth projective schemes.
\end{dfn}
Let $e \in X$ be a point. Take $\pi_0= e \times X$ and $\pi_{2d} = X \times e$ where $d:= \dim X$ and $X$ is irreducible. Write ${\mathfrak{m}}^{0}_{\sim}(X):= (X, \pi_{0},0)$ and ${\mathfrak{m}}^{2d}_{\sim}(X):= (X, \pi_{2d}, 0)$. We can show that
\begin{equation*}
{\mathfrak{m}}^{2d}_{\sim}(X) \simeq (\Spec k, \id, -d)
\end{equation*}
and use this to show the fundamental fact
\begin{equation*}
{\bf L} \simeq {\mathfrak{m}}^{2}_{\sim}(\Proj^1, \Proj^1 \times e, 0)
\end{equation*}
where $\Proj^1$ is the projective line over $k$ and $e$ a point in $\Proj^1$.
\par
The category of (virtual) pure motives $\mot_{\sim}(k)$ has the following properties.
\begin{itemize}
\item{$\mot_{\sim}(k)$ is an additive category. That is, $\Hom_{\mot_{\sim}(k)}(M,N)$ are abelian groups and $\oplus$ exists: $M \oplus N:= (X \amalg Y, p \amalg q, m)$ for $\dim X = \dim Y = m$.}
\item{$\mot_{\sim}(k)$ is a pseudoabelian category; it is additive with a well-defined image of $p$.}
\item{There is a tensor structure on $\mot_{\sim}(k)$: $$M \otimes N:= (X \times Y, p \times q, m+n).$$}
\item{There is a multiplicative structure on $\mot_{\sim}(k)$: $$m_X: \mathfrak{m}_{\sim}(X) \otimes \mathfrak{m}_{\sim}(X) \simeq \mathfrak{m}_{\sim}(X \times X) \stackrel{\mathfrak{m}_{\sim}(\Delta)}{\lra} \mathfrak{m}_{\sim}(X).$$}
\end{itemize}
Let $k$ be a field and $\overline{k}$ be its algebraic closure. Fix an equivalence relation $\sim$ on the cycles of $X$. Let $M$ be a virtual pure motives of a smooth projective variety $X$, i.e, $M = \mathfrak{m}_{\sim} (X)$.
\begin{dfn}[Realization of a pure motive]
Define the realization of a motive $M$ as the functor
\[
\mot_{\sim}(k) \stackrel{\real}{\lra} \grvect 
{\mathfrak m}_{\sim} (X) \mapsto H^{*}(X_{\overline{k}}, \Q)
\]
for a Weil cohomology functor $H^{*}(-, \Q)$. We say that a motive can be realized if there exists at least one Weil cohomology for which the functor $\real$ is exact and faithful for all  adequate relations $\sim$. 
\end{dfn}
The reader may set $H^{*}(X_{\overline{k}},\Q)$ as algebraic de Rham cohomology as  to fix ideas. 
\par
Grothendieck's original conception of motives was such that the following diagram of functors commute \emph{for all adequate equivalence relations $\sim$ and all Weil cohomologies $H^{*}(X, \Q)$}:
\begin{equation*}\begin{CD}
\mot_{\sim}(k) @= \mot_{\sim}(k) \\
@A{\mathfrak{m}_{\sim}(-)}AA           @VV{\real(-)}V\\
\smproj^{op} @>{H^{*}(-, \Q)}>> \grvect
\end{CD} \end{equation*}
There are three categories of motives of particular interest to us: the category of \emph{Grothendieck motives} when the equivalence relation is homological equivalence $\mhom(k)$, the category of \emph{Chow motives} when the equivalence relation is homological equivalence $\mchow(k)$ and the category of \emph{numerical motives} when the equivalence relation is numerical equivalence $\mnum(k)$ . We have the following ``fundamental theorem of pure motives":
\begin{thm} \label{tannakamot}
$\mhom(k)$ can be realized for all Weil cohomologies and is neutral Tannakian with the fiber functor $\real$.
\end{thm}
This is the reason why number theorists often use the term ``motive of $X$" when they mean ``the $l$-adic cohomology of $X$ with $\Q_l$-coffecients and a continuous $G_k$ action on it". (The action of $G_k$ is through the motivic Galois group of $X$, cf. section \ref{motgalgp}.)
\par
Recall the yoga of Tannakian formalism of section \ref{tannaka}. Deligne defines the dimension of an object in a rigid tensor category $\textbf{C}$ in the following way:
\begin{dfn}[dimension]
Let $\point$ be the unit object in $\textbf{C}$. Let $M \in \obj({\textbf{C}})$ be an arbitrary object. Let $f \in \End M$ and trace of $f$ on $M$ is defined by the composition of maps
$$\point \stackrel{\delta}{\lra} \widehat{M} \otimes M \stackrel{t}{\lra} M \otimes \widehat{M} \stackrel{\ev}{\lra} \point$$
and denote as $\tr f_M$. (The map $t$ exchanges the factors in $\widehat{M} \otimes M$.) The dimension of $M$ is $ \underline{\dim} M := \tr 1_M$.
\end{dfn}
A result of Deligne show that dimension is always positive: $\textbf{C}$ rigid tensor $\iff$ for all $M \in \obj ({\textbf{C}})$, $\underline{\dim} M \in \N$.
\par
Furthermore for $\mhom(k)$, Deligne shows that this definition of dimension coincides (through the Weil conjectures) with our usual notion of cohomological dimension: 
$$\underline{\dim} M = \dim H^{*}(X, \Q) := \oplus_i \dim H^{i}(X, \Q)$$ 
where $H^{i}(X, \Q)$ is a realization of the motive $M$. The Standard Conjectures imply that $\dim M$ is independent of the choice of the Weil cohomology.

\begin{thm}[Jannsen]\label{jannsen}
The category $\mnum(k)$ is semisimple and abelian.
\end{thm}
\subsection{Motivic Galois groups}\label{motgalgp} The Tannakian formalism along with theorem \ref{tannakamot} gives
\begin{equation*}
\repk(\Aut^{\otimes} H^{*}_B(-, \Q)) \stackrel{\sim}{\lra} \mot_{\hom}(k) \stackrel{\real(-)}{\lra}   \grvect \circlearrowleft G_{\textup{mot}}\\
\end{equation*}
where $G_{\textup{mot}}:= \Aut^{\otimes} H^{*}_B(-, \Q)$ is the \emph{motivic Galois group}. It  acts on the image of the realization functor and is a proalgebraic group. (It is proreductive assuming Jannsen's theorem \ref{jannsen} and the Conjecture D(X) \ref{fundamental}.) 
\par
If $\mot_{\sim}(k)$ is generated by $\mathfrak{m}_{\sim} (\Spec E)$ where $E$ is a finite extension of $k$ then $G_\textup{mot} = G_k$. If $\mot_{\sim}(k)$ is generated by the Lefschetz motive $\textbf{L}$, then $G_\textup{mot} = \Gm$.

\section{Mixed motives}\label{mixedmotives}

Pure motives are motives of smooth projective varieties. In most physical applications of the theory of motives, most notably in the case of the motives of hypersurfaces associated to Feynman graphs in perturbative quantum field theory, working with smooth projective varieties is too restrictive. In fact, one knows that the projective hypersurfaces obtained from the parametric form of Feynman integrals are typically singular, and this already leaves the world of pure motives. Moreover, since the parametric Feynman integral (see the appendix) is computed over a domain of integration with boundary, what is involved from the motivic point of view is really a relative cohomology of the hypersurface complement, relative to a normal crossings divisor that contains the boundary of the domain of integration. This is a second reason why it is mixed motives and not pure motives that are involved, since these are the natural environment where long exact cohomology sequences and relative cohomologies live. 
Unfortunately, from the mathematical point of view, the theory of mixed motives is far more complicated than that of pure motives. At present, one only has a triangulated category of mixed motives, with various equivalent constructions due to Voevodsky, Levine, and Hanamura.
Only in very special cases, such as mixed Tate motives over a number field, it is possible to construct and abelian category. While we are not going to give any details on the construction of the triangulated category of mixed motives, we recall the relevant cohomological
properties and some properties of their analytic counterpart, mixed Hodge structures. We also recall some preliminary notions about triangulated categories to aid the reader in her further study.

\subsection{Derived and triangulated categories}\footnote{Because of the rather dry and technical nature of the precise definitions, I have tried to give a more general idea about triangulated and derived categories; the following are a collection of concepts as oppossed to a list of formal definitions and properties.} This material is taken from Dimca \cite{dimca}. Let $\abelian$ be an abelian category. Let $C(\abelian)$ be the category of complexes in $\abelian$. (The reader unfamiliar with homological algebra should immediately set $\abelian$ to be the category of $R$-modules $\modu$.) $C(\abelian)$ contains three important full subcategories $C^{\bullet}(\abelian)$:
\begin{enumerate}
\item{$C^{+}(\abelian)$ of complexes bounded on the left: 
\begin{equation*}
\cdots \lra 0 \lra \cdots \lra A^{-1} \lra A^{0} \lra \cdots 
\end{equation*}}
\item{$C^{-}(\abelian)$ of complexes bounded on the right: 
\begin{equation*}
\cdots \lra A^{0} \lra A^{1} \lra \cdots \lra 0 \lra \cdots
\end{equation*}}
\item{$C^{b}(\abelian)$ the full subcategory of complexes bounded both on the right and the left.}
\end{enumerate}
Let $X^{\bullet}, Y^{\bullet}$ be two complexes in $C^{\bullet}(\abelian)$. Call the morphism $u: X^{\bullet} \ra Y^{\bullet}$ a \emph{quasi-isomorphism} if at the level of cohomology $H^{k}(u): H^{k}(X^{\bullet}) \ra H^{k}(Y^{\bullet})$ is an isomorphism for all $k$. 
\par
Define a \emph{shift automorphism} on complexes $T: C^{\bullet} \ra C^{\bullet}$ as
\begin{eqnarray*}
(X[n])^r := X^{n+r},\\
d^{s}_{T(X^{\bullet})} := - d^{s+1}_{X^{\bullet}}
\end{eqnarray*}
for the complex
\begin{equation*}
A^{\bullet}: \cdots \lra A^{m-1} \stackrel{d^{m-1}}{\lra} A^{m} \stackrel{d^{m}}{\lra} A^{m+1} \stackrel{d^{m+1}}{\lra} \cdots
\end{equation*}
Let $u: X^\bullet \rightarrow Y^\bullet$ be a morphism of complexes in $C^\bullet(\abelian)$. The {\em mapping cone} of the morphism $u$ is the complex in $C^\bullet(\abelian)$ given by
$$C^\bullet_u = Y^\bullet \oplus (X^\bullet[1]).$$
This gives rise to the {\em standard triangle} for a morphism $u$
$$T_u: X^\bullet \stackrel{u}{\lra} Y^\bullet \stackrel{q}{\lra} C^\bullet_u \stackrel{p}{\lra} X^\bullet[1],$$
where $q$ (resp. $p$) is the inclusion (resp. projection) morphisms. The standard triangle gives rise to long exact sequences in cohomology.
\par
An important construction is that of a {\em homotopic} category\footnote{The reader is invited to compare this with the definition of a Grothendieck {\em group} in \ref{Kuniv}.}. This is an additive category $K^\bullet(\abelian)$ with
\begin{itemize}
\item{$\obj(K^\bullet(\abelian)) = \obj(C^\bullet(\abelian))$,}
\item{$\Hom_{K^\bullet(\abelian)}(X^\bullet, Y^\bullet) = {\Hom(X^\bullet, Y^\bullet)}/\sim$ where $\sim$ is homotopy equivalence: $u \sim v \implies H^k(u) = H^k(v)$.}
\end{itemize}
A family of triangles in $K^\bullet(\abelian)$ is {\em distinguished} if they are isomorphic to a standard triangle for some morphism $u$.
\par
In a homotopic category $K^\bullet(\abelian)$, distinguished triangles satisfy a list of four properties referred to as TR1--TR4 in the literature. We will not repeat them here (see, for example, proposition 1.2.4 of \cite{dimca}). It suffices to say that TR1--TR4 guarantees nice functorial properties of distinguished triangles, compatible with homotopy. 
\par
A {\em triangulated category} is an additive category $\mathcal{A}$ with a shift self-equivalence $T$, with $X[1]=TX$ and with a collection of distinguished triangles $\mathcal{T}$ that satisfy TR1--TR4 in the original definition due to Verdier. A {\em derived category} $D^\bullet(\abelian)$ of an abelian category $\abelian$ is a triangulated category obtained from $K^\bullet(\abelian)$ by localization with respect to the multiplicative system of quasi-isomorphisms in $K^\bullet(\abelian)$.

\subsection{Bloch--Ogus cohomology}\label{bocoh}
The material for this section is taken from the seminal paper of Bloch--Ogus \cite{blochogus}. Bloch-Ogus cohomology is the mixed motives counterpart of Weil cohomology in the pure motives case, and as such, is a universal cohomology theory for schemes of a much general type.
\par
Let $\sch$ be the category of schemes of finite type over the field $k$. By  ${\sch}^{*}$ I mean the category with 
\begin{itemize}
\item{Objects: $Y \hra X$ closed immersions in $X \in \obj(\sch)$ and}
\item{Morphisms: ${{\Hom}_{{\sch}^{*}}}((Y \hra X), (Y' \hra X'))$ being the following commutative cartesian squares:
\begin{equation*}\begin{CD}
Y @>{\subset}>> X \\
@V{f_Y}VV           @VV{f_X}V\\
Y' @>{\subset}>> X'
\end{CD} \end{equation*}
}
\end{itemize}
\begin{dfn}[cohomology with supports]\label{cohsupp}
A \emph{twisted cohomology theory with supports} is a sequence of contravariant functors 
\begin{eqnarray*}
H^{i}: {\sch}^{*} &\lra& \abgps\\
(Y \hra X) &\mapsto& \bigoplus_i H^{i}_Y (X,n),\\
\end{eqnarray*}
satisfying the following:
\begin{enumerate}
\item{For $Z \subseteq Y \subseteq X$, there is a long exact sequence
$$\cdots \lra H^{i}_Z(X,n) \lra H^{i}_Y(X,n) \lra H^{i}_{Y\setminus Z} (X\setminus Z,n) \lra H^{i+1}_Z(X,n) \lra \cdots.$$}
\item{Let \begin{eqnarray*}
f: (Y \hra X) & \lra & (Y' \hra X'), \\
g: (Z \hra X) & \lra & (Z' \hra X')
\end{eqnarray*} 
and $k: (Y \setminus Z \hra X \setminus Z) \ra (Y' \setminus Z' \hra X' \setminus Z')$ be the induced arrow. Let $h: (Z \hra X) \ra (Z' \hra X')$. Then the following diagram commutes:
\begin{equation*}\begin{CD}
@>>> H^{i}_Z(X,n) @>>> H^{i}_Y (X,n) @>>> H^{i}_{Y \setminus Z}(X \setminus Z,n) @>>> H^{i+1}_{Z}(X,n) @>>> \\
& & @A{H^{*}(h)}AA    @A{H^{*}(f)}AA     @A{H^{*}(k)}AA                      @A{H^{*}(g)}AA \\
@>>> H^{i}_{Z'}(X',n) @>>> H^{i}_{Y'} (X',n) @>>> H^{i}_{Y' \setminus Z'}(X' \setminus Z',n) @>>> H^{i+1}_{Z'}(X',n) @>>> \\
\end{CD} \end{equation*}
}
\item{[Excision] Let $(Z \hra X) \in \obj({\sch}^{*})$ and $(U \hra X)$ be open in $X$ containing $Z$. Then
$$H^{i}_Z(X,n) \stackrel{\sim}{\lra} H^{i}_Z(U,n).$$}
\end{enumerate}
\end{dfn}
Dually we have
\begin{dfn}\label{homsupp}
[Homology with supports] A \emph{twisted homology with supports} is a sequence of covariant functors
$$H_{*}: {\sch}_{*} \lra \abgps$$
where the category ${\sch}_{*}$ has objects of $\sch$ and the morphisms are proper morphisms of $\sch$ satisfying:
\begin{enumerate}
\item{$H_{*}$ is a presheaf in \'etale topology. If $\alpha: X' \ra X$ is an \'etale morphism there exists morphisms$\alpha^{*}: H_i(X,n) \ra H_i(X',n)$.}
\item{Let $\alpha: Y' \ra Y$ and $\beta: X' \ra X$ be \'etale. Let $f: X \ra Y$ and $g: X' \ra Y'$ be proper and consider the following cartesian square:
\begin{equation*}\begin{CD}
X' @>{\beta}>> X \\
@V{g}VV           @VV{f}V\\
Y' @>{\alpha}>> X'
\end{CD} \end{equation*}
}
Then the following square commutes:
\begin{equation*}\begin{CD}
H_i(X',n) @<{\beta^{*}}<< H_i(X,n) \\
@V{H_i(g,n)}VV           @VV{H_i(f,n)}V\\
H_i(Y',n) @<{\alpha^{*}}<< H_i(Y,n) 
\end{CD} \end{equation*}
\item{Let $i: Y \hra X$ be a closed immersion and $\alpha: X \setminus Y \hra X$ be an open immersion. Then there exists a long exact sequence
$$\cdots \lra H_{i}(Y,n) \stackrel{i^{*}}{\lra} H_{i}(X,n) \stackrel{\alpha^{*}}{\lra} H_{i}(X\setminus Y,n) \lra H_{i-1}(Y,n) \lra \cdots.$$}
\item{Let $f: X' \ra X$ be proper and $Z = f(Z')$ for $Z \hra X$. Let $\alpha: X' \setminus f^{-1}(Z) \hra X' \setminus Z'$. Then the following square commutes:
\begin{equation*}\begin{CD}
@>>> H_i(Z',n) @>>> H_{i} (X',n) @>>> H_{i}(X' \setminus Z',n) @>>> H_{i-1}(Z',n) @>>> \\
& & @V{f_{*}}VV    @V{f_{*}}VV     @V{f_{*}\alpha^{*}}VV                    @V{f_{*}}VV \\
@>>> H_{i}(Z,n) @>>> H_{i} (X,n) @>>> H_{i}(X \setminus Z,n) @>>> H_{i-1}(Z,n) @>>> \\
\end{CD} \end{equation*}
}
\end{enumerate}
\end{dfn}

\begin{dfn}[Poincar\'e duality with supports] \emph{A Poincar\'e duality theory for schemes of finite type with supports} is a twisted cohomology theory $H^{*}$ with
\begin{enumerate}
\item{For all $Y \hra X \in \obj({\sch}^{*})$ there is a pairing
$$H_i(X,n) \times H^{j}_Y(X,n) \lra H_{i+j}(Y, m+n).$$}
\item{If $Y \hra X \in \obj({\sch}^{*})$ and 
$(\beta \hra \alpha): (Y' \hra X') \lra (Y \hra X)$ an \'etale morphism in ${\sch}^{*}$, then for $a \in H^{j}_Y (X,n)$ and $z \in H_i(X,m)$,
$$\alpha^{*} (a) \cap \alpha^{*}(Z) = \beta^{*}(a \cap z).$$}
\item{[Projection] Let $f: (Y_1 \hra X_1) \lra (Y_2 \hra X_2)$ proper. Then for $a \in H^{i}_{Y_2}(X_2, n)$ and $z \in H_i(X_1, m)$,
$$H^{i}(f_X) (z) \cap a = H_{i}(f_Y)(z \cap H^{i}(f)(a)).$$}
\item{[Fundamental class] Let $X \in \obj(\sch)$ be irreducible and of dimension $d$. There exists a global section $\eta_X$ of $H_{2d}(X,d)$ such if $\alpha: X' \lra X$ is \'etale, then $\alpha^{*}\eta_X = \eta_{X'}$.}
\item{[Poincar\'e duality] Let $X \in \obj(\sch)$ be smooth and of dimension $d$. Let $Y \hra X$ be a closed immersion. Then
$$H^{2d-i}_{Y}(X, d-n) \stackrel{\cap{\eta_X}}{\lra} H_i(Y,n)$$
is an isomorphism.}
\end{enumerate}
\end{dfn}
The main theorem of \cite{blochogus} is
\begin{thm}\label{PDblochogus}
Given a Poincar\'e duality theory, an \'etale morphism $f_X$ and the commutative square
\begin{equation*}\begin{CD}
Z' @>>> X' \\
@V{f_Z}VV           @VV{f_X}V\\
Z @>{\subset}>> X
\end{CD} \end{equation*}
the following diagram commutes:
\begin{equation*}\begin{CD}
H^{i}_{Z'}(X',n) @>{\cap \eta_{X'}}>> H_{2d-i}(Z', d-n)\\
@A{H^{*}(f)}AA           @AA{f^{*}_{Z}}A\\
H^{i}_{Z}(X,n) @>{\cap \eta_{X}}>> H_{2d-i}(Z, d-n)
\end{CD} \end{equation*}
\end{thm}

\subsection{Hodge structures} \label{hodgestr} Let us quickly review the basic definitions of Hodge theory.
\begin{dfn}[Pure Hodge structure]\label{hodgepure}
A pure Hodge structure of weight $m$ on a finite dimensional vector space $V$ is a decreasing filtration
\begin{equation*}
\cdots \subset F^{p+1} V_\C \subset F^{p}V_\C \subset \cdots
\end{equation*}
\end{dfn}
with $V_\C:= V \otimes_\R \C$ and satisfying the \emph{Hodge decomposition}
\begin{equation*}
V_\C = \oplus_{p+q = m} V^{p,q},
\end{equation*}
where $V^{p,q} = F^{p} \cap \overline{F^{p}V_\C}$ and $\overline{}$ denotes the conjugate filtration.
\end{dfn}
Hodge structures form a category $\hs$ with morphisms of vector spaces compatible with the filtration $F$ making up the hom set. The category $\hs$ is a tensor category with the evident tensor product. Furthermore, by taking formal differences of Hodge structures $[H] - [H']$, we obtain the Grothendieck ring of Hodge structures $K(\hs)$. 
\par
Let $A \subset \R$. An $A$-mixed Hodge structure consists of the following data:
\begin{enumerate}
\item{An $A$-module of finite type $V_A$.}
\item{An increasing filtration called the \emph{weight filtration}
\begin{equation*}
\cdots \subset W_n \subset W_{n+1} \subset \cdots
\end{equation*}
of $A \otimes \Q$-module $V_A \otimes \Q$.}
\item{A decreasing filtration called the \emph{Hodge filtration}
\begin{equation*}
\cdots F^{p+1} V_\C \subset F^{p}V_\C \subset \cdots
\end{equation*}
where $V_\C \otimes V_A \otimes \C$.}
\item{A \emph{graded weight $j$ factor} $\gr_j^{W}(V_A) := (W_j/W_{j-1}) \otimes \C$ with a pure Hodge struture induced by the filtration $F$ and $\overline{F}$ on $V_\C$.}
\end{enumerate}
Mixed Hodge structures too form a category $\mhs$ with morphisms $V_A \ra V_{A'}$ $A$-module homomorphisms that are compatible with the Hodge and weight filtrations. The following theorem is of great importance:
\begin{thm}[Deligne]
The category $\mhs$ is abelian with kernels and cokernels with induced filtrations.
\end{thm}
\subsection{Mixed Tate motives over a number field} Mixed Tate motives can be defined as 
the triangulated subcategory of the triangulated category of mixed motives generated by the 
Tate objects $\Q(n)$. Over a number field, however, it is possible to obtain a nicer category
in the following way.

Let $k$ be a number field. Let $\Q(1)$ be the pure Tate motive. The category of mixed Tate motives over $k$, denoted as $\mtmot_k$ is constructed in the following way. Consider the simple objects $\Q(n)$ and assume that $\Q(a)$ and $\Q(b)$ are isomorphic for $a \neq b$ and that any simple object of $\mtmot_k$ is isomorphic to some $\Q(\cdot)$. Also consider the groups
\begin{equation*}
\Ext^{i}_{\mtmot_k}(\Q(0), \Q(n))
\end{equation*}
and assume that they vanish for $i > 1$. An important result of Borel identifies these extensions:
\begin{thm}[Borel]
$\Ext^{1}_{\mtmot_k}(\Q(0), \Q(n)) \simeq K_{2n-1}(k) \otimes \Q$
where $K_{2n-1}(k)$ is the Quillen $K$-theory of the field $k$. 
\end{thm}
Now $K_{2n-1}(k) \otimes \Q = H_{\bullet} (GL(n,k))$, the linear group with entries in $k$, so the reader may think of the Quillen $K$-theory in this case in terms of this homological identification. We have the following fundamental theorem:
\begin{thm}[Deligne--Goncharov]
The category $\mtmot_k$ for a number field $k$ is a Tannakian category with objects $\Q(n)$ and the extensions described above. Furthermore, the Hodge realization functor $\mtmot_k \ra \mhs$ is exact and faithful.
\end{thm} 

\section{Motivic measures and zeta functions}\label{motzeta}
The material for this section is based on \cite{denefloeser} and \cite{looijenga}.
\subsection{Overview of basics} The \emph{generalized Euler characteristic} $\chi$ associates to each object in $\quasiproj$ an element in a fixed commutative ring $R$ such that $\chi(X) = \chi(Y)$ for $X \simeq Y$ and $\chi(X) = \chi(Y) + \chi(X \setminus Y)$ for a closed subvariety $Y \subset X$. The product in $R$ is given by the fibered product of varieties: $\chi(X \times Y) = \chi(X)\chi(Y)$. In fact the Euler characteristic is the canonical example of a generalized Euler characteristic. 
\par
Konstevich's original motivation behind inventing motivic measures was to prove the following theorem
\begin{thm}[Kontsevich]
Let $X$ and $Y$ be two birationally equivalent Calabi-Yau manifolds. Then $X$ and $Y$ have the same Hodge numbers.
\end{thm}
These types of questions are of enormous importance for duality questions in string theory, namely in mirror symmetry. (There is a beautiful theory around this called the \emph{geometric McKay correspondence} which I completely omit from the discussion.)
\par
An example, following the proof of this theorem of Kontsevich, of a generalized Euler characteristic, comes from Hodge numbers of a complex manifold $X$. Recall the definition of mixed Hodge structures. For a Hodge structure $V$, we can define the class of $V$ in $K(\hs)$ in terms of the graded $m$-factors in the weight filtration: $[V] := \sum_m [\gr^W_m (V)]$. Define the \emph{Hodge-Deligne polynomial} as
\begin{equation*}
\chi_h(X) := \sum_i[H^{i}_c(X, \Q)] \in K(\hs)
\end{equation*}
where $H^{i}_c(X,\Q)$ is the $i$-th cohomology of $X$ with compact support. If $Y \subset X$ is locally closed, then the Hodge characteristic is compatible with the exact Gysin sequence:
\begin{equation*}
\cdots \lra H^r_c(X \setminus Y, \Q) \lra H^r_c(X, \Q) \lra H^r_c(Y, \Q) \lra H^{r+1}_c(X \setminus Y, \Q) \lra \cdots.
\end{equation*} 
That is 
\begin{equation*}
\chi_h(X) = \chi_h(Y) + \chi_h(X \setminus Y).
\end{equation*}
\begin{rem}
For $X$ the affine line $\A^1_k$, $H^r_c(\A^1_k, \Q) = 0$ for all $ r \neq 2$ and $H^2_c(\A^1_k, \Q)$ is one-dimensional of Hodge type $(1,1)$. So $\chi_h(\A^1_k)$ is invertible. This is the Lefschetz motive $\lef$. 
\end{rem}
We have the following two ring homomorphisms
\begin{itemize}
\item{[Hodge characteristic] 
\begin{eqnarray*}
\chi_h: K(\hs) \ra \Z[u, u^{-1}, v, v^{-1}]\\
H^{i}(X,\Q) \mapsto \sum_{p+q=i} \dim(H^{p,q}(X, \Q)) u^p v^q
\end{eqnarray*}}
\item{[Weight characteristic]
\begin{eqnarray*} 
\chi_{wt}: K(\hs) \ra \Z[w,w^{-1}]\\
u,v \mapsto w.
\end{eqnarray*}}
\end{itemize}
Here $H^{p,q}(X, \Q)$ is the $(p,q)$-th piece of $H^i_c(X, \Q)$. Evaluating the weight characteristic at 1 gives us the topological Euler characteristic. 
\subsection{Equivariant Grothendieck ring}\label{subs:equivgroth}
Denote by $\hat{\mu}$ the projective limit of the affine scheme of roots of unity $\mu_n$:
\begin{equation*}
\hat{\mu}:= \limproj_n \mu_n = \limproj_n \Spec k[x]/(x^n -1)
\end{equation*}
Let $X$ be an $S$-variety. A \emph{good $\mu_n$-action on $X$} is a group action $\mu_n \times X \ra X$ such that each orbit is contained in an affine subvariety of $X$. (A \emph{good $\hat{\mu}$-action} factors through a good $\mu_n$-action for some $n$.)
\begin{dfn}[equivariant Grothendieck ring]
The equivariant Grothendieck ring of varieties $K^{\hat{\mu}}(\relvar_{X_0})$ is a an abelian group generated by $[X, \hat{\mu}]_S$ for $X$ an $S$-variety with good $\hat{\mu}$-action with the relations 
\begin{enumerate}
\item{$[X, \hat{\mu}]_S = [Y, \hat{\mu}]_S$
for $X \simeq Y$ as $S$-varieties with good $\hat{\mu}$-action.}
\item{$[X,\hat{\mu}]_S = [Y,\hat{\mu}]_S + [X \setminus Y,\hat{\mu}]_S$ if $Y$ is closed in $X$ and the $\hat{\mu}$-action on $Y$ induced by $\hat{\mu}$-action on $X$.}
\item{[monodromy] $[X \times V,\hat{\mu}]_S = [\A^n_k,\hat{\mu}]_S$ where $V$ is the $n$-dimensional affine space with a good $\hat{\mu}$-action and $\A^n_k$ is the affine space with a trivial $\hat{\mu}$-action.}
\end{enumerate}
\end{dfn}
\begin{dfn}[equivariant Euler characteristic] The \emph{equivariant Euler characteristic} is a ring homomorphism
\begin{eqnarray*}
\chi_{\textrm{top}}(-, \alpha): K^{\hat{\mu}}(\relvar_{X_0})[\lef^{-1}] \lra  Z\\
(-)  \mapsto \sum_{q \geq 0} (-1)^q \dim H^q(-, \C)_\alpha
\end{eqnarray*}
where $H^q(-, \C)_\alpha \subset H^{*}(-, \C)$ on which there is a good $\hat{\mu}$-action through multiplication by $\alpha$.
\end{dfn}

\subsection{Arc spaces} Let $k$ be a field of characteristic zero and let $X$ be a variety over $k$. For all natural numbers $n$, we define the \emph{arc space} $\scrl_n(X)$ as an algebraic variety over $k$ whose $k$-rational points are $K[t]/ t^{n+1}$-rational points for some $k \subset K$. We set $\mathcal{L}(X) = \limproj_n \scrl_n (X)$. Note that $\scrl_0 (X) = X$ and $\scrl_1(X) = TX$, the (Zariski) tangent space of $X$. We call the $K$-rational points of $\scrl_n(X)$ $K$-arcs of $X$ or arcs for short. There are the structure morphisms 
\begin{eqnarray*}
\pi_n: \scrl(X) \ra \scrl_n(X),\\
\pi^m_n: \scrl_m(X) \ra \scrl_n(X).
\end{eqnarray*} 
The \emph{origin} of an arc $\gamma$ is $\pi_0(\gamma)$.  
\par
A ``cleaner" (but equivalent) defintion of the space of arcs is the following:
\begin{dfn}[Space of arcs] \label{dfn:arc}
Let $X$ be a variety over $k$ (which, I remind the reader is a separated scheme of finite type over $k$.) Denote by $\scrl(X)$ the \emph{scheme of germs of arcs on $X$}. It is a defined as a scheme over $k$ such that, for any extension $k \subset K$ there is a natural bijection
\begin{equation*}
\scrl(X)(K) \simeq \Hom_{\sch} (\Spec k[[t]], K)
\end{equation*}
The scheme $\scrl(X) := \limproj_n \scrl_n(X)$ in the category of schemes $\scrl_n(X)$ representing the functor
\begin{equation*}
R \mapsto \Hom_{\sch}(\Spec R[t]/t^{n+1}, X)
\end{equation*}
defined on the category of $K$-algebras. 
\end{dfn}
Not much is known about the space $\scrl(X)$. An important result is the following:
\begin{propn}[Kolchin]
Let $X$ be an integral scheme. Then $\scrl(X)$ is irreducible.
\end{propn} 
\subsection{The Nash Problem} Arcs were first introduced by Nash in connection with singularities: let $P$ be a singular point on $X$ and set $\scrl_{\{P\}}(X):= \pi^{-1}_0(P)$ of arcs with origin $P$. He studied the space ${\mathcal{N}}_{\{P\}}(X) \subset \scrl_{\{P\}}(X)$ of arcs not contained in the singular locus $\sing X$. Let me briefly sketch Nash's beautiful idea, following Loeser's Trieste lectures \cite{loeser}.
\par 
Let $Y \stackrel{\rho}{\ra} X$ be a resolution of singularities of a scheme $X$. Recall that this means: $Y$ is smooth, $\rho$ is proper and $\rho$ induces an isomorphism $Y \setminus \rho^{-1}(\sing X) \simeq X \setminus \sing X$. We say that the resolution of singularities is \emph{divisorial}  if the locus $E$ where $\rho$ is \emph{not} a local isomorphism (called the \emph{exceptional set}) is a divisor in $Y$. In such cases, we call the locus an \emph{exceptional divisor}. Let $Y' \stackrel{\rho'}{\ra} X$ be another proper birational morphism (and assume that $X$ is normal) and let $p$ be the generic point of $E$. We say that \emph{$E$ appears in $\rho'$} if ${\rho'}^{-1} \circ \rho$ is a local isomorphism at $p$. The exceptional divisor $E$ is called an \emph{essential component} of a resolution of $X$ if $E$ appears in every divisorial resolution of $X$. Denote the set of essential components as ${\mathcal{C}}_{\{P\}}(X)$.
\begin{thm}[Nash]
The mapping ${\mathcal{C}}_{\{P\}}(X) \stackrel{\nu}{\lra} {\mathcal{N}}_{\{P\}}(X)$ is injective.
\end{thm}
Basically, what this theorem tells is that every irreducible component of $\scrl(X)$ through a singularity $P$ corresponds to an exceptional divisor that occurs on every resolution. The \emph{Nash Problem} asks
\begin{question}
For what $X$ is the map $\nu$ a bijection?
\end{question}
An explicit nonexample was provided by Ishii-Koll\'ar
when $X$ is a 4-dimensional hypersurface singularity $\{(x_1,x_2, x_3, x_4, x_5) \in \A^5_k| x^3_1 + x^3_2 + x^3_3 + x^3_4 + x^6_5 =0\}$ which has 1 irreducible family of arcs but 2 essential components. (The base $k$ is obviously not allowed to be of characteristic 2 or 3.)

\subsection{Motivic zeta functions} Let $t$ be a \emph{fixed} coordinate on $\A^1_k$ and let $n \geq 1$ be an integer. A morphism $f: X \ra \A^1_k$ induces a morphism $f_n: \scrl_n(X) \ra \scrl_n(\A^1_k)$. Any $\alpha \in \scrl(\A^1_k)$ (resp. $\alpha \in \scrl_n(\A^1_k)$) gives a power series $\alpha(t) \in K[[t]]$) (resp. $\alpha(t) \in K[[t]]/t^{n+1}$) by definition. Define a map $\ord_t: \scrl(\A^{1}_k) \ra \Z_{\geq 0} \cup \{\infty\}$ as $\ord_t(\alpha) := {\textrm{max}}_{t^e | \alpha(t)} \{e\}$. 
\begin{dfn}
$\mathcal{X}_n := \{ \phi \in \scrl_n(X)| \ord_t f_n(\phi) = n \}$.
\end{dfn}
The variety $\mathcal{X}_n$ is a locally closed subvariety of $\scrl_n(X)$. Let $X_0$ be a variety obtained by setting $f=0$. The variety $\mathcal{X}_n$ is an $X_0$-variety through the structure maps $\pi^n_0$. Define the morphism 
\begin{equation*}
\overline{f}_n: \scrx_n \ra \Gm \textrm{ given by }\phi \mapsto \textrm{coefficient of }t_n \textrm{ in } f_n(\phi)
\end{equation*}
and set $\scrx_{n,1} := {\overline{f}}^{-1}_n (1)$. The following characterization of $\mathcal{X}_n$ allows us to pass on to \'etale settings ``without reductions at bad primes". 
\begin{propn}
As $\Gm \times X_0$-variety, $\scrx_n$ is a quotient of $\scrx_{n,1} \times \Gm$ by the $\mu_n$-action $a(\phi,b) = (a\phi, a^{-1}b)$.
\end{propn}
Recall the definition of the equivariant Grothendieck ring $K^{\hat{\mu}}(\relvar_{X_0})$ of section \ref{subs:equivgroth}.
\begin{dfn}[Denef-Loeser, Looijenga]\label{dfn:motzeta}
The \emph{motivic zeta function} of a morphism $f: X \ra \A^{1}_k$ over $K^{\hat{\mu}}(\relvar_{X_0})[\lef^{-1}]$ is defined by the sum
\begin{equation*}
Z(T) = \sum_{n \geq 1} [\scrx_{n+1}, \hat{\mu}]_{X_0} \lef^{-nd} T^{n}
\end{equation*}
where $[-, \hat{\mu}]_{X_0}$ is a class in $K^{\hat{\mu}}(\relvar_{X_0})$ and $d = \dim \scrx_{n+1}$.
\end{dfn}
\begin{thm}[Denef-Loeser]
The motivic zeta function $Z(T)$ in definition \ref{dfn:motzeta} is a rational function.
\end{thm}

\appendix
\section{Motivic ideas in physics (by M.Marcolli)}

The theory of motives was introduced to mathematicians by Grothendieck as a universal cohomology theory for algebraic varieties. At present, while a lot of progress happened since its origin, it is still a field in very rapid development, with a wealth of intriguing conjectures still unsolved and new unexpected applications being found. Among
these, motives recently made their appearance in the world of theoretical physics, a latecomer with respect to more traditional mathematical tools adopted by the physics community, but one that is likely to lead to a wider range of future applications. We attempt here to convince our audience that, in various forms, some of the main ingredients of the theory of motives have already penetrated the horizon of the high-energy physics community. Roughly speaking, the theory of motives encompasses an {\em analytic} side, which is based on Hodge theory and mixed Hodge structures, a {\em geometric} side based on algebraic varieties and algebraic cycles, and an {\em algebraic} side coming from algebraic $K$-theory. This appendix will recall briefly some contexts familiar to theoretical physics where these notions have come play an important role. We will try to focus not only on the setting of perturbative scalar quantum field theories and the parametric representation of Feynman integrals, which is one of the main interactions between physics and motives of interest today, but also on other occurrences of some of the main motivic ideas in other aspects of theoretical physics. 

\subsection{The dawn of motives in physics}
One can trace the origin of motivic ideas in physics to the theory of electromagnetism and the ``cohomological" interpretation of the Maxwell equations in terms of differential forms. In fact, electromagnetism is the historic origin of Hodge theory, which in turn
is on of the essential building blocks of the theory of motives. Naturally, the current ideas on mixed Hodge structures are a long distance away from the Maxwell equations, but as we discuss briefly below, ideas related to Hodge theory continued to play a role in physics beyond electromagnetism, most notably in the more recent context of mirror symmetry. On the other hand, periods of algebraic varieties, another important aspect of the theory of motives, have made repeated appearances in the context of theoretical physics, from their role in residues of Feynman graphs, which is being extensively investigated today, to the modeling of spectral density functions in condensed matter physics. 

\subsection{Hodge theory: from electromagnetism to mirror symmetry}
An important way in which some motivic ideas make contact with the world of physics is through Hodge theory. In fact, the very origin of Hodge theory is in the formulation of Maxwell's equations of electromagnetism. In terms of electric and magnetic fields, these are
usually written in the form
\begin{equation}\label{Maxwell}
\left\{\begin{array}{rl}
\nabla \times E =& \frac{\partial B}{\partial t} \\[2mm]
\nabla \times B =& -\frac{\partial E}{\partial t} \\[2mm]
{\rm div}\, E =& 0 \\[2mm]
{\rm div}\, B =& 0,
\end{array} \right.
\end{equation}
with $E=(E_x,E_y,E_z)$ and $B=(B_x,B_y,B_z)$. One can better assemble the electric and magnetic field in a single entity, described by a 2-form 
$$ F =(E_x dx+E_y dy + E_z dz)\wedge dt+ B_x dy\wedge dz -B_y dx\wedge
dz + B_z dx\wedge dy. $$
The advantage of this formulation is that it transforms the Maxwell equations \eqref{Maxwell} into the much more appealing and conceptually simple form
\begin{equation}\label{Maxwell2}
\left\{ \begin{array}{rl} dF & =  0 \\[2mm] d^* F & = 0,
\end{array}\right.
\end{equation}
where $d$ is the deRham differential with adjoint $d^*$. See the beautiful paper of Raoul Bott \cite{Bott} for a more detailed discussion of these and other examples of geometrization of physics.

This formulation of the Maxwell equations leads immediately to considerations on the Hodge decomposition of differential forms 
$\omega = d\alpha + d^*\beta + \gamma$, with $\gamma$ harmonic,
satisfying $d\gamma=d^*\gamma =0$ as above. The identification of deRham cohomology with harminic forms, in turn, relates then to the Hodge decomposition of the cohomology of compact K\"ahler manifolds 
$$ H^k(X,\C) = \oplus_{p+q=k} H^{p,q}(X), $$
which is the origin of the notion of Hodge structures as the ``analytic side'' of the theory of motives of algebraic varieties. 

In this more refined form, Hodge theory made a spectacular return into the world of theoretical physics with the discovery of mirror symmetry. This originated with the observation that certain Calabi-Yau threefolds, which serve as compactified dimensions in string theory, form mirror pairs when their $\sigma$-models determine the same 
superconformal field theory. In particular, it was observed that two such mirror varieties $X$ and $X^\prime$, while geometrically very different, have a very simple ``mirror relation'' between their Hodge numbers
\begin{equation}\label{mirrors}
h^{p,q}(X) = h^{3-p,q}(X^\prime),
\end{equation}
where $h^{p,q}(X)=\dim H^{p,q}(X)$. From this initial observation the field of mirror symmetry rapidly developed into an extremely interesting and elaborate branch of contemporary mathematical physics, where several sophisticated ideas from algebraic geometry and arithmetic geometry, including various aspects of the theory of motives such as algebraic cycles, came to play an increasingly important role. 

\subsection{Algebraic cycles and homological mirror symmetry}

The main way in which algebraic cycles came to play a role in the more recent developments of homological mirror symmetry (see \cite{Kaz}) is through the fact that they define objects in the bounded derived category $D^b(X)$ of coherent sheaves. For a variety of general type that is realized as a complete intersection in a toric variety, homological mirror symmetry predicts that the derived category $D^b(X)$ can be realized as a subcategory of the Fukaya--Seidel category $F(Y,W)$ of a mirror Landau--Ginzburg model specified by a symplectic manifold $Y$ with a superpotential $W$. As shown in \cite{Kaz}, using this approach one can transform questions about algebraic cycles into corresponding questions in the mirror environment, typically in terms of the information associated to the singularities of $W$. This point of view may prove very useful from the motivic viewpoint. In fact, one of the main source of difficulties in the theory of motives is the fact that questions about algebraic cycles are extremely difficult to approach, which is the reason why Grothendieck's``standard conjectures'' on motives remain unsolved. One can, potentially, at least in some cases, employ the homological mirror symmetry viepoint to relate such questions, as in \cite{Kaz}, to constructions involving vanishing cycles and singularity theory in the mirror Landau-Ginzburg model.

\subsection{Algebraic varieties and periods: Fermi surfaces}

Periods are a very interesting class of complex numbers, which are in general not algebraic, but which can be obtained via a geometric procedure where all the objects involved are themselves algebraic. More precisely, these are numbers that are obtained by
integrating an algebraic differential form over a cycle in an algebraic variety. For the purpose of our setting here, we can assume that the varieties themselves are defined over $\Q$ (or more generally a finite extension, a number field) and so are the algebraic differential forms, while the domain of integration is a cycle defined again over the same field. One also allows for the case where the domain of integration has boundary, in which case it is defined by inequalities, as a semi-algebraic set, again over the same field, and
in this case the integration of a differential form correspond to a pairing with a relative cohomology group. We'll see this in more explicit examples later. There are relations between periods coming from change of variables in the integrals that compute them and from
the Stokes formula. The properties, known and conjectural, of the algebra of periods are analyzed in detail in the paper of Kontsevich--Zagier \cite{kz}. The type of numbers that one obtains as periods can be used as numerical signatures of the motivic complexity of the algebraic variety. For example, multiple zeta values are the tell-tale sign of the presence of a particular class of motives, the mixed Tate motive, about which more later.  

A first beautiful example of the role of periods of algebraic varieties arises in the context of Fermi surfaces in condensed matter, see \cite{GiKnoTru1}, \cite{GiKnoTru2}, \cite{Peters}. One considers electrons moving in a lattice, in the independent electron approximation, where one can describe the dynamics as a single particle model with an effective potential. The energy levels of the system are obtained from the Schr\"odinger equation
\begin{equation}\label{Schrod1}
(-\Delta + V) \psi = \lambda \psi, \ \ \ \text{ with } \ \ \
\psi(x+\gamma)= e^{i\langle k, \gamma\rangle} \psi(x),
\end{equation}
for a fixed $k\in \R^d$ and for all $\gamma \in \Gamma$, the lattice of ions. The $j$-th eigenvalue $\lambda_j(k)$, for given crystal momentum $k$, defines the $j$-th band  in the spectrum.

The Fermi surface separates in the space of crystal momenta the occupied and non-occupied states at zero temperature. One typically replaces the spectral problem \eqref{Schrod1} with a discretized version, with the Laplacian replaced by a Harper operator (discrete
Laplacian) on a lattice, $H=\sum_j (T_j + T_j^*)$, summed over the generators of the lattice, with $T_j$ the shift operators,
\begin{equation}\label{Schrod2}
(-H + V -\lambda) \psi =0, \ \ \ \text{ with } T_j^{a_j} \psi = \xi_j \psi,
\end{equation}
with $T_j^{a_j}$ the shifts corresponding to the lattice $\Gamma = \oplus_{j=1}^d \Z a_j$. In these geometric terms, one then describes the Fermi surface in terms of the Bloch variety
\begin{equation}\label{BlochVar}
B(V)=\{ (\xi,\lambda)\in (\C^*)^d \times  \C\,|\, \exists \psi \text{
solution of } \eqref{Schrod2} \}.
\end{equation}
The Fermi surface $F_\lambda$ is obtained from the Bloch variety as the fiber $F_\lambda = \pi^{-1}(\lambda)$ of the projection $\pi: B(V)\to \C$. This is an algebraic variety, from which one can read physical properties of the system. One of the main results of \cite{GiKnoTru1} is the fact that the {\em density of states} $\rho(\lambda)$ of the system can be computed as a {\em period} of this algebraic variety in the form
\begin{equation}\label{rhostates}
\frac{d\rho}{d\lambda}=\frac{1}{(2\pi)^d a_1\cdots a_d}
\int_{F_\lambda \cap (S^1)^d} \omega_\lambda,
\end{equation}
where the domain of integration are the real points of the Fermi surface $F_\lambda$ and the algebraic differential form is defined by the
relation
$$ \pi^*(d\lambda)\wedge \omega_\lambda = \frac{d\xi_1}{\xi_1} \wedge \cdots \wedge \frac{d\xi_d}{\xi_d}. $$
Although the algebro-geometric properties of the Bloch varieties and Fermi surfaces have been studied extensively in \cite{GiKnoTru1}, there has been so far no explicitly motivic analysis of this family of varieties and of the corresponding periods. A notable exception is the paper of Steinstra \cite{Stien}, which gives a motivic interpretation of the period in terms of Deligne cohomology and Mahler measures, in the case of zero potential. One can expect that a more extensive analysis of the motivic properties of Bloch varieties and Fermi surfaces will lead to interesting results, in the same vein and with similar techniques as those adopted in the study of the motivic aspects of Feynman integrals described here below.

\subsection{The parametric form of Feynman integrals}

Another example of the role in physics of periods of algebraic varieties, which is presently being widely investigated, is that of parametric Feynman integrals of perturbative scalar quantum field theories. 

In perturbative quantum field theory, to a scalar field theory with Euclidean Lagrangian of the form
\begin{equation}\label{Lagr}
{\mathcal L}(\phi) = \int \left( \frac{1}{2} (\partial_\mu \phi)^2 
+ \frac{m^2}{2}\phi^2 + {\mathcal P}(\phi) \right) dv ,
\end{equation}
with the interaction term given by a polynomial ${\mathcal P}(\phi)$, one assigns a formal perturbative expansion
$$ \sum_\Gamma \frac{U(\Gamma,p_1,\ldots,p_N)}{\# Aut(\Gamma)} $$
parameterized by Feynman graphs $\Gamma$. These are finite graphs with internal edges connecting pairs of vertices and external edges attached to a single vertex and carrying extrernal momenta $p_i$. The vertices are constrained to have valences equal to the
powers of the monomials in ${\mathcal P}(\phi)$. The contribution $U(\Gamma,p_1,\ldots,p_N)$ is dictated by the Feynman rules of the theory, and can be expressed as a (typically divergent) finite dimensional integral. In the parametric form (or $\alpha$-parametrization) these integrals can be written in the form
\begin{equation}\label{UGammaPPsi}
U(\Gamma,p_1,\ldots,p_N) =
\frac{\Gamma(n-\frac{D\ell}{2})}{(4\pi)^{D\ell/2}} 
\int_{\sigma_n}
\frac{P_\Gamma(t,p)^{-n+D\ell/2}\omega_n}{\Psi_\Gamma(t)^{-n +D(\ell+1)/2}} ,
\end{equation}
where $n=\# E_{int}(\Gamma)$ and $\ell = b_1(\Gamma)$. The domain of integration is the simplex $\sigma_n = \{ t \in \R_+^n\,|\, \sum_i
t_i =1 \}$.
The Kirchhoff polynomial (also called first Symanzik polynomial) $\Psi_\Gamma(t)$ is given by 
\begin{equation}\label{SpanTrees}
\Psi_\Gamma(t) = \sum_{T\subset \Gamma} \prod_{e\notin E(T)} t_e,
\end{equation}
where the sum is over all the spanning trees (forests) $T$ of the graph $\Gamma$ and for
each spanning tree the product is over all edges of $\Gamma$ that are not in the tree. The second Symanzik polynomial $P_\Gamma(t,p)$ is similarly defined in terms of cut sets instead of spanning trees, and it depends explicitly on the external momenta of the graph (see \cite{BjDr2} \S 18). The polynomial $\Psi_\Gamma(t)$ is homogeneous of degree $\ell=b_1(\Gamma)$, while, in the massless case, the polynomial $P_\Gamma(t,p)$ is homogeneous of degree $\ell+1$.

After removing the divergent Gamma factor from \eqref{UGammaPPsi}, one is left with the residue of the Feynman integral, which has then the form of a period, in the sense that it is the integration of an algebraic differential form on a semi-algebraic set in the algebraic variety given by the complement of the hypersurface defined by the vanishing of the graph polynomial in the demonimator of the integrand of \eqref{UGammaPPsi}. Saying that the integral is a period, of course, ignores the important issue of divergences: in fact, even after removing the Gamma factor, the remaning integral can still contain divergences coming from the intersections of the domain of integration with the hypersurface. Modulo this issue, which needs to be addressed via a suitable regularization and renormalization
procedure, one is considering a period whose value contains information on the motivic nature of the relative cohomology
\begin{equation}\label{relcohom}
H^{n-1}({\mathbb P}^{n-1}\smallsetminus X_\Gamma, \Sigma_n \smallsetminus
(X_\Gamma \cap \Sigma_n)),
\end{equation}
where $\Sigma_n$ is the divisor of coordinate hyperplanes in ${\mathbb P}^{n-1}$, which contains the boundary of the domain of integration $\partial \sigma_n \subset \Sigma_n$. The frequent appearance of multiple zeta values in Feynman integral computations (see \cite{BroKr}) suggests that the cohomology \eqref{relcohom} may be a realization of a mixed Tate motive. However, \cite{bb1} shows that the graph hypersurfaces $X_\Gamma$ themselves can be arbitrarily complex in motivic terms: their classes $[X_\Gamma]$ span the Grothendieck ring of varieties. Understanding precise conditions on the graphs under which the Feynman integral gives a period of a mixed Tate motive is presently a very interesting open question in the field.

\subsection{Motivic Galois groups in physics}

A manifestation of motivic Galois groups in physics arises in the context of the Connes--Kreimer theory of perturbative renormalization. In \cite{cm1} it was shown that the counterterms in the BPHZ renormalization procedure, described as in Connes--Kreimer
in terms of Birkhoff factorization in the affine group scheme dual to the Hopf algebra of Feynman graphs, can be equivalently formulated in terms of solutions to a certain class of differential systems with irregular singularities. This is obtained by writing the terms in the Birkhoff factorization as time ordered exponentials, and then using the fact that
$$ {\rm T} e^{\int_a^b\,\alpha(t)\,dt} :=\,1+\, \sum_{n=1}^\infty \int_{a\leq
s_1\leq \cdots\leq
s_n\leq b} \,\alpha(s_1)\cdots\,\alpha(s_n)\,\,  ds_1\cdots ds_n $$
is the value $g(b)$ at $b$ of the unique solution $g(t)\in G$ with value $g(a)=1$ of the differential equation $ dg(t)=\,g(t)\,\alpha(t)\,dt$.

The type of singularities are specified by physical conditions, such as the independence of
the counterterms on the mass scale. These conditions are expressed geometrically through the notion of $G$-valued {\em equisingular connections} on a principal $\C^*$-bundle
$B$ over a disk $\Delta$, where $G$ is the pro-unipotent Lie group of characters of the Connes--Kreimer Hopf algebra of Feynman graphs. In physical terms, the disk $\Delta$ gives the complexified dimension of dimensional regularization of the Feynman integrals, while the
fiber over $z\in \Delta$ consists of the $\mu^z$, with the parameter $\mu$ giving the energy scale of the renormalization group flow. The {\em equisingularity} condition is the property that such a connection $\omega$ is $\C^*$-invariant and that its restrictions to sections of the principal bundle that agree at $0\in \Delta$ are mutually equivalent, in the sense that they are related by a gauge transformation by a $G$-valued $\C^*$-invariant map regular in $B$, hence they have the same type of (irregular) singularity at the origin. The classification of equivalence classes of these differential systems can be done via the Riemann--Hilbert correspondence and differential Galois theory. This means assembling these data, in the form of a category of ``flat equisingular vector bundles'', which one can show
(\cite{cm1}) is a Tannakian category. The Tannakian formalism then identifies it with the category of finite dimensional linear representations of a Tannakian Galois group $U^*=U\rtimes {\mathbb G}_m$, where $U$ is pro-unipotent with Lie algebra the free graded Lie algebra with one generator $e_{-n}$ in each degree $n\in \N$. The group $U^*$ is identified (non-canonically) with the motivic Galois group of mixed Tate motives over the cyclotomic ring $\Z[e^{2\pi i/N}]$, for $N=3$ or $N=4$, localized at $N$.

\subsection{Algebraic K-theory and conformal field theory}

Algebraic $K$-theory, which one can think of as the more algebraic manifestation of the theory of motives, may appear at first to be the part that is more remote from physics. On the contrary, it has come to play an important role, for example in the context of conformal
field theory. We mention here very briefly some of the recent developments in this direction, while we refer the reader to \cite{Nahm} for a more informative overview.

Recall that the dilogarithm is defined as 
$$ Li_2(z)=\int_z^0 \frac{\log (1-t)}{t}\, dt = 
\sum_{n=1}^\infty \frac{z^n}{n^2}. $$
It satisfies the functional equation $Li_2(z)+Li_2(1-z)=Li_2(1)-\log(z) \log(1-z)$, where $Li_2(1)=\zeta(2)$, for $\zeta(s)$ the Riemann zeta function. A variant is given by the Rogers dilogarithm $$ L(x)=Li_2(x)+\frac 12 \log(x)\log(1-x).$$ See \cite{Zagier} for a detailed survey on the dilogarithm and its properties.

There is a relation between the torsion elements in the algebraic $K$-theory group $K_3(\C)$ and rational conformally invariant quantum field theories in two dimensions, see \cite{Nahm}. There is, in fact, a map, given by the dilogarithm, from torsion elements in the Bloch group (closely related to the algebraic $K$-theory) to the central charges and scaling
dimensions of the conformal field theories.

This correspondence arises by considering sums of the form
\begin{equation}\label{sumAq}
\sum_{m\in \N^r} \frac{q^{Q(m)}}{(q)_m},
\end{equation}
where $(q)_m=(q)_{m_1}\cdots (q)_{m_r}$, with $(q)_{m_i}= (1-q)(1-q^2)\cdots (1-q^{m_i})$ and $$ Q(m)=m^t A m/2 + b m + h$$ has rational coefficients. Such sums are naturally obtained from considerations involving the partition function of a bosonic rational CFT. In particular, \eqref{sumAq} can define a modular function only if all the solutions of the equation
\begin{equation}\label{Blochgr}
\sum_j A_{ij} \log(x_j)= \log (1-x_i)
\end{equation}
determine elements of finite order in an extension $\hat B(\C)$ of the Bloch group, which accounts for the fact that the logarithm is multi-valued. The Rogers dilogarithm gives a natural group homomorphism $(2\pi i)^2 L: \hat B(\C) \to \C/\Z$, which takes values in $\Q/\Z$ on the torsion elements. These values give the conformal dimensions of the fields in the theory.

\bibliographystyle{amsalpha}

\begin{thebibliography}{A}


\bibitem{paomat} Paolo Aluffi, Matilde Marcolli \textit{Feynman motives of banana graphs} (2008) arXiv:0801.1690v2 [hep-th]

\bibitem{paomatfeynrules} Paolo Aluffi, Matilde Marcolli \textit{Algebro-geometric Feynman rules} (2008) arXiv:0811.2514v1 [hep-th]

\bibitem{paomatfeyndet} Paolo Aluffi, Matilde Marcolli \textit{Parametric Feynman integrals and determinant hypersurfaces} (2009) arXiv:0901.2107v1 [math.AG]


\bibitem{andre} Yves Andr\'e textit{Une introduction aux motifs (motifs purs, motifs mixtes, p\'eriodes)} Soci\'et\'e Math\'ematique de France, Paris, 2004.

\bibitem{bb1} Prakash Belkale, Patrick Brosnan \textit{Matroids, motives and a conjecture of Kontsevich} Duke Math. J. \textbf{116} (2003) 147--188.


\bibitem{bittner} Franziska Bittner \textit{The universal Euler characteristic for varieties of characteristic 	zero} Compos. Math. \textbf{140} 4 (2004) 1011--1032.


\bibitem{BjDr2} James D. Bjorken, Sidney D. Drell \textit{Relativistic quantum fields}, McGraw-Hill Book Co. New York, 1965.


\bibitem{sp1} Spencer Bloch \textit{Motives associated to graphs} Jpn. J. Math. \textbf{2} 1 (2007) 165--196.


\bibitem{blochsum} Spencer Bloch \textit{Motives associated to sums of graphs} (2008) arXiv:0810.1313v1 [math.AG]


\bibitem{blochogus} Spencer Bloch, Arthur Ogus \textit{Gersten conjectures and the homology of schemes} Annals. Sci. ENS \textbf{7} (1974).


\bibitem{BEK} Spencer Bloch, H{\'e}l{\`e}ne Esnault, Dirk Kreimer \textit{On motives associated to graph polynomials} Comm. Math. Phys. \textbf{267} 1 (2006) 181--225.


\bibitem{blochkreimer} Spencer Bloch, Dirk Kreimer \textit{Mixed Hodge structures and renormalization in physics} Commun.Num.Theor.Phys. \textbf{2} 4 (2008) 637--718.


\bibitem{bognerweinzierl} Christian Bogner, Stefan Weinzierl \textit{Periods and Feynman integrals} (2007) arXiv:0711.4863v2 [hep-th]


\bibitem{Bott} Raoul Bott \textit{On some recent interactions between mathematics and physics} Canad. Math. Bull. \textbf{28} 2 (1985) 129--164.


\bibitem{BroKr} D.J. Broadhurst, D. Kreimer \textit{Association of multiple zeta values with positive knots via Feynman diagrams up to {$9$} loops} Phys. Lett. B. \textbf{393} 3--4 (1997) 403--412.


\bibitem{ck1} Alain Connes, Dirk Kreimer \textit{Renormalization in quantum field theory and the Riemann-Hilbert problem I: The Hopf algebra structure of graphs and main theorem} Comm. Math. Phys. \textbf{210} (2000) 249--273.


\bibitem{ck2}{article} Alain Connes, Dirk Kreimer \textit{Renormalization in quantum field theory and the Riemann-Hilbert problem I: The $\beta$-function, diffeomorphisms and the renormalization group} Comm. Math. Phys. \textbf{216} (2001) 215--241.


\bibitem{cm1} Alain Connes, Matilde Marcolli \textit{Renormalization, the Riemann-Hilbert correspondence, and motivic Galois theory} Frontiers in number theory, physics, and geometry (II) Springer, Berlin (2007) 617--713.
   

\bibitem{cmbook} Alain Connes, Matilde Marcolli \textit{Noncommutative geometry, quantum fields and motives} American Mathematical Society, Providence, RI, 2008.


\bibitem{deligneweil} Pierre Deligne \textit{La conjecture de Weil. I} Inst. Hautes \'Etudes Sci. Publ. Math. \textbf{43} (1974) 273--307.

		
\bibitem{delignetannaka} Pierre Deligne \textit{Cat\'egories tannakiennes} The Grothendieck Festschrift, Vol. II, Birkh\"auser Boston (1990) 111--195.


\bibitem{denefloeser} Jan Denef, Francois Loeser \textit{Geometry of arc spaces on algebraic varieties} (2000) arXiv:0006050v1 [math.AG]

\bibitem{dimca} Alexandru Dimca \textit{Sheaves in topology} Springer-Verlag, New York, 2003.

\bibitem{doryn} Dzmitry Doryn \textit{Cohomology of graph hypersurfaces associated to certain Feynman graphs} Doctoral thesis, Universit\"at Duisburg-Essen (2008) arXiv:0811.0402v1 [math.AG]


\bibitem{fulton} William Fulton \textit{Intersection theory} Springer-Verlag, New York, 1998.

\bibitem{gelfandmanin} S.I. Gelfand, Y.I. Manin \textit{Methods of homological algebra} Springer-Verlag, Berlin, 2003.

 
\bibitem{gillet} H. Gillet. C. Soul{\'e} \textit{Descent, motives and $K$-theory} J. Reine Angew. Math. \textbf{478} (1996) 127--176.


\bibitem{GiKnoTru1} D. Gieseker, H. Kn{\"o}rrer, E. Trubowitz \textit{The geometry of algebraic Fermi curves} Academic Press Inc. Boston MA, 1993.


\bibitem{GiKnoTru2} D. Gieseker, H. Kn{\"o}rrer, E. Trubowitz \textit{An overview of the geometry of algebraic Fermi curves} Contemp. Math. \textbf{116} Amer. Math. Soc. Providence RI. (1991) 19--46.
		

\bibitem{grifharr} Phillip Griffiths, Joseph Harris \textit{Principles of algebraic geometry} John Wiley \& Sons Inc. New York, 1994.

	
\bibitem{SGA1}{book} Alexander Grothendieck \textit{Rev\^etements \'etales et groupe fondamental. Fasc. I: Expos\'es 1 \`a 5} S\'eminaire de G\'eom\'etrie Alg\'ebrique I \textbf{1960/61} Institut des Hautes \'Etudes Scientifiques, Paris, 1963.
   
				
\bibitem{standconj} Alexander Grothendieck \textit{Standard conjectures on algebraic cycles}, Algebraic Geometry (Internat. Colloq., Tata Inst. Fund. Res., Bombay, 1968) Oxford Univ, London (1969) 193--199.


\bibitem{hales} Thomas C. Hales \textit{What is motivic measure?} Bull. Amer. Math. Soc. (N.S.) \textbf{42} 2 (2005) 119--135.
	
				
\bibitem{hart} Robin Hartshorne \textit{Algebraic geometry} Springer-Verllag, New York, 1977.

\bibitem{looijenga} Eduard Looijenga \textit{Motivic measures} (2000) arXiv:0006220v2 [math.AG]


\bibitem{loeser} Francois Loeser \textit{Motivic integration and McKay correspondence} Lectures at ``Trieste School and Conference on Intersection Theory and Moduli" (2002)


\bibitem{bible} \textit{Motives} Proceedings of the AMS-IMS-SIAM Joint Summer Research Conference held at the University of Washington, Seattle, Washington, July 20--August 2, 1991, American Mathematical Society, Providence RI, 1994.
   
	
\bibitem{Kaz} Ludmil Katzarkov  \textit{Homological Mirror Symmetry and Algebraic Cycles} Homological Mirror Symmetry, Lecture Notes in Physics \textbf{757} Springer Verlag (2008) 1--28.
					
\bibitem{kz} Maxim Kontsevich, Don Zagier \textit{Periods} Mathematics unlimited--2001 and beyond, Springer, Berlin (2001), 771--808.

\bibitem{lang} Serge Lang \textit{Algebra} Springer-Verlag, New York, 2002.


\bibitem{larsen} Michael Larsen \textit{Motivic measures and stable birational geometry} Mosc. Math. J. \textbf{3} 1 (2003) 85--95.
		

\bibitem{matilde} Matilde Marcolli \textit{Motivic renormalization and singularities} (2008) arXiv:0804.4824v2 [math-ph]


\bibitem{matildeecm} Matilde Marcolli \textit{Feynman integrals and motives} Pleanry lecture at European Congress of Mathematicians 2008 (2008).


\bibitem{marrej} Matilde Marcolli, Abhijnan Rej \textit{Supermanifolds from Feynman graphs} J. Phys. A \textbf{41} (2008) 315402--315423.


\bibitem{maninmot} Yuri I. Manin \textit{Correspondences, motifs and monoidal transformations} (Russian) Mat. Sb. (N.S.) \textbf{77 (119)} (1968) 475--507.

		
\bibitem{manin1} Yuri I. Manin \textit{Lectures on zeta functions and motives (according to Deninger and Kurokawa)} Ast\'erisque \textbf{228} (1995) 121--163.
  

\bibitem{mansuper} Yuri I. Manin \textit{Gauge field theory and complex geometry} Springer-Verlag, Berlin, 1997.

	
\bibitem{murre} Jacob P. Murre \textit{Lectures on motives} Transcendental aspects of algebraic cycles, London Math. Soc. Lecture Note Ser. \textbf{313} Cambridge Univ. Press (2004) 123--170.

  
\bibitem{Nahm} Werner Nahm \textit{Conformal field theory and torsion elements of the Bloch group} Frontiers in number theory, physics, and geometry. {II} Springer, Berlin (2007) 67--132.


\bibitem{Peters} Chris Peters \textit{Algebraic {F}ermi curves (after Gieseker, Trubowitz and Kn\"orrer)} S{\'e}minaire Bourbaki, Vol. 1989/90, Ast\'erisque \textbf{189--190} (1990) Exp.\ No.\ 723, 239--258.
	

\bibitem{poonen} Bjorn Poonen \textit{The Grothendieck ring of varieties is not a domain} Math. Res. Lett. \textbf{9} 4 (2002) 493--497.


\bibitem{rosenberg} Jonathan Rosenberg \textit{Algebraic $K$-theory and its applications} Springer-Verlag, New York, 1994.

\bibitem{sav} Neantro Saavedra Rivano \textit{Cat\'egories Tannakiennes} Lecture Notes in Mathematics \textbf{265} Springer-Verlag, Berlin, 1972.


\bibitem{neeraja} Neeraja Sahasrabudhe \textit{Grothendieck ring of varieties} Master's Thesis, Universit\'e Bordeaux 1 (2007).


\bibitem{schorev} A.J. Scholl \textit{Classical motives} Motives Seattle WA 1991, Proc. Sympos. Pure Math. \textbf{55} Amer. Math. Soc., Providence, RI (1994) 163--187.

 
\bibitem{shin} Sug Woo Shin \textit{Grothendieck function-sheaf correspondence} Lecture notes from the Harvard seminar on geometric class field theory (2005).
	

\bibitem{Stien} Jan Steinstra \textit{Motives from diffraction} Algebraic cycles and motives. Vol. 2, London Math. Soc. Lecture Note Ser. \textbf{344} Cambridge Univ. Press (2007) 340--359.

		
\bibitem{weibel} Charles Weibel \textit{The $K$-book: An introduction to algebraic $K$-theory} in preparation, chapters I--IV available at http://www.math.rutgers.edu/$\sim$weibel/Kbook.html


\bibitem{Zagier} Don Zagier \textit{The dilogarithm function} Frontiers in number theory, physics, and geometry. II, Springer, Berlin (2007) 3--65.

\end{thebibliography}

\end{document}